\documentclass[prb,aps,twocolumn,groupaddress,longbibliography]{revtex4-1}
\usepackage{latexsym}
\usepackage{lineno}
\usepackage{hyperref}
\usepackage{url}
\usepackage{graphicx}
\usepackage{verbatim}
\usepackage{multirow}
\usepackage{amsmath}
\newcommand{\beq}{\begin{eqnarray}}
\newcommand{\eeq}{\end{eqnarray}}
\usepackage{mathrsfs}
\usepackage{float}
\usepackage[usenames, dvipsnames]{color}
\usepackage{mathtools}
\usepackage{slashed}
\usepackage{physics}	
\usepackage{graphicx}   
\usepackage{epstopdf}
\usepackage{subfigure}  
\usepackage{hyperref}   
\usepackage{bbold}
\usepackage{wasysym}
\usepackage{amssymb}
\usepackage{feynmp}
\usepackage[symbol]{footmisc}

\usepackage[latin1]{inputenc}
\usepackage{tikz}
\usetikzlibrary{decorations.pathmorphing}
\usetikzlibrary{shapes.misc}
\tikzset{cross/.style={cross out, draw=black, minimum size=8*(#1-\pgflinewidth), inner sep=0pt, outer sep=0pt},
cross/.default={1pt}}
\usetikzlibrary{patterns,math}
\hypersetup{
     colorlinks=true,
     linkcolor=blue,
     filecolor=blue,
     citecolor = red,      
     urlcolor=blue,
     }
     
     \definecolor{bluemathematica}{rgb}{0.368417, 0.506779, 0.709798} 
      \definecolor{orangemathematica}{rgb}{0.880722, 0.611041, 0.142051} 
     
\begin{document}

\title{Superconducting dome in ferroelectric-type materials from soft mode instability}

\author{Chandan Setty}
\thanks{csetty@ufl.edu}
\affiliation{Department of Physics and Astronomy, Rice Center for Quantum Materials, Rice University, Houston, Texas 77005, USA}
\author{Matteo Baggioli}
\thanks{b.matteo@sjtu.edu.cn}
\affiliation{Wilczek Quantum Center, School of Physics and Astronomy, Shanghai Jiao Tong University, Shanghai 200240, China \& Shanghai Research Center for Quantum Sciences, Shanghai 201315.}
\author{Alessio Zaccone}
\thanks{alessio.zaccone@unimi.it}
\affiliation{Department of Physics ``A. Pontremoli", University of Milan, via Celoria 16, 20133 Milan, Italy.}
\affiliation{Cavendish Laboratory, University of Cambridge, JJ Thomson Avenue, CB30HE Cambridge, U.K.}

\begin{abstract}
We present a minimal theory of superconductivity enhancement in ferroelectric-type materials. Simple expressions for the optical mode responsible for the soft mode transition are assumed. A key role is played by the anharmonic phonon damping which is modulated by an external control parameter (electron doping or mechanical strain) causing the appearance of the soft mode. It is shown that the enhancement in the superconducting critical temperature $T_{c}$ upon approaching the ferroelectric transition from either side is due to the Stokes electron-phonon scattering processes promoted by strong phonon damping effects. 
\end{abstract}
\maketitle

\section{Introduction} 
Although it was originally thought that ferroelectricity  and ferroelectric-type crystals cannot support superconductivity\cite{Matthias1955},  connections between the two phenomena were explored\cite{Blount1965, Appel1966, Appel1969} early on, and a possible origin of superconductivity was understood in terms of dielectric screening\cite{Ginzburg1987} on the pairing mechanism or the role of ferroelectric instabilities\cite{Buttner1989}. Since then, a number of systems have been discovered where superconductivity may coexist with a quantum critical ferroelectric transition\cite{Binnig,Lonzarich2020}, and where the presence of the soft mode appears to enhance the superconducting critical temperature $T_{c}$. Most notably, this is the case of strontium titanate, SrTiO$_{3}$, for which several conventional and unconventional mechanisms have been discussed\cite{Appel1969,Pfeiffer1967,Spaldin} since its early discovery\cite{Cohen1964}. 
Even though the unusual conditions due to extreme dilution of electrons make the pairing mechanism in SrTiO$_{3}$ still a highly debated topic, a basic BCS framework can be applied to extract qualitative pairing trends. \cite{Behnia2013,Chubukov2016}

More recently, enhanced superconductivity domes in the proximity of soft mode instabilities, induced by electron doping~\cite{Ma2021} and mechanical strain~\cite{Balatsky}, have been observed. The role of structural instability in promoting superconductivity is suggested by the recent finding that plastic mechanical deformations appear to further enhance superconductivity possibly due to self-organized dislocation networks~\cite{Hameed2021}.
{Various theoretical explanations\cite{Gastiasoro,Ruhman,Spaldin,Balatsky,Koreans} have been proposed to understand such effects in these ferroelectric-type materials, although a complete picture of the mechanism of $T_c$ enhancement near the instabilities is unclear.}

The role of structural instability, and its influence on superconductivity, similar to what happens in ferroelectrics\cite{Shapiro2019}, is also a highly debated topic in the context of high-temperature superconductors, both cuprates\cite{Phillips,Shapiro2021} and the recently discovered hydrides under high pressure\cite{Eremets2020,Chen2020,Rawal2021}, and, in general, in strongly coupled superconductors\cite{Sadovskii}.
For example, it is well known that, before the discovery of cuprates, high-$T_c$ superconductivity was sought in A15 intermetallic compounds which exhibit soft-mode structural instability\cite{Testardi}. {While soft mode instabilities and phonon softening have been observed in several cuprates\cite{LeTacon2014,Uchiyama2004,Dastuto2008,Reznik2008,Lee2021} always below the $T_c$ \cite{Pepin2021}, their role in promoting superconductivity remains uncertain\cite{Chen2020}.}

In all the relevant experimental systems~\cite{Spaldin,Balatsky,Ma2021,Valentinis}  the anharmonic phonon damping plays a key role\cite{Chen2020}, which is reflected in gigantic values of the Gr\"{u}neisen parameter as recently observed in Ref.\cite{Balatsky}. If the soft mode is induced by electron doping, the damping arises from the enhanced electron-phonon coupling; if, instead, it is induced by mechanical strain, the damping is increased by the sheer interatomic potential anharmonicity as the strain pushes atoms away from the harmonic part of the bonding wells.

Hence, it is imperative to understand the effect of the soft mode on superconductivity by directly considering the effect of the growing anharmonic phonon damping accompanying the soft mode instability.
From the perspective of effective theories, we proposed a minimal theoretical model based on the anharmonically extended BCS theory~\cite{Setty2020,Setty2021}, which includes the effect of anharmonic damping of phonons mediating the superconductivity.\\
In this work, we demonstrate that the superconducting dome in ferroelectric-type materials, or, more generally, systems with soft-mode structural instabilities, originates from the enhancement of the superconducting critical temperature $T_c$ due to anharmonic phonon damping that peaks at the instability point. An analogue of this effect, in the absence of soft mode instabilities, was recently predicted theoretically~\cite{Setty2020} and was found to be consistent with recent experimental observations~\cite{Shibauchi}.

\section{Theoretical framework} 
\subsection{Soft mode description}
We start by assuming an optical phonon undergoing a ferroelectric-type soft mode transition as a function of a generic external control parameter $n$, which could be e.g. mechanical strain or electron doping.
The external control parameter acts on the complex optical dispersion relation by affecting the phonon damping, i.e. by increasing it up to the point where the phonon energy vanishes and the soft mode is created.

In general, at low enough values of the frequency $\omega$ and the wave-vector $k$, within the so-called \textit{hydrodynamic expansion}, the optical phonon dispersion relation satisfies \cite{Kosevich}
\begin{equation}
-\omega^2 + \omega_{0}^2 - i\, \omega \,\Gamma = 0\label{dd}
\end{equation}
where $\omega_{0}$ is the phonon frequency (already renormalized by anharmonicity) and $\Gamma$ is the anharmonic phonon damping, which is quantitatively related to the Gr\"{u}neisen coefficient\cite{Balatsky} via the Klemens relation\cite{Klemens}. Here, higher corrections in the wave-vector are ignored for simplicity.
In ferroelectric-type materials the optical phonon that undergoes the softening is typically the tranvserse optical (TO) phonon\cite{Kittel}, and therefore its coupling to electronic degrees of freedom is weak or negligible due to the vanishing of the dot product\cite{Ruhman,Gorkov}. Recent work\cite{Gastiasoro2020,Gastiasoro2021} (see also Ref.\cite{Koreans}) showed that, in reality, a significant coupling of the soft TO phonon to electrons can be induced by Rashba effects due to spin-orbit coupling. Other mechanisms have been also proposed which can enhance the coupling\cite{Rischau,Coleman2021,Feigelman,Ngai}.
In our effective model, we remain agnostic as to the origin of the coupling of the soft optical phonon to the electrons and we instead focus on the role of anharmonic damping connected with the softening.

The presence of the damping $\Gamma >0$ causes both a renormalization of the natural (bare) phonon energy, the real part of the frequency $\omega$ appearing in Eq.\eqref{dd}, which becomes:
\begin{equation}
    \mathrm{Re}(\omega)\,=\,\frac{1}{2}\,\sqrt{4\,\omega_0^2\,-\,\Gamma^2}<\omega_0\,,\label{realpart}
\end{equation}
as well as the finite lifetime $\tau = (\Gamma/2)^{-1}$ (its imaginary part). Both these effects can be computed for a given lattice structure and interatomic interactions via Self-Consistent Phonon (SCP) theory~\cite{Baowen,Tadano1,Tadano2} by considering three and four-phonon processes, i.e. by accounting for anharmonicity of the lattice.
Here, we shall assume that the damping $\Gamma$ is also a function of an additional external degree of freedom, which could be mechanical strain or electron doping. In the former case, varying the mechanical strain causes the atoms to sample more anharmonic parts of the interaction potential. In the latter case, electronic doping changes the electron-phonon damping which also contributes to $\Gamma$. 

We consider an effective description and we remain agnostic about the exact nature of the microscopic processes that change $\Gamma$ as our goal is to have a generic model to study how variations in $\Gamma$ induced by a generic external parameter $n$ lead to changes in the superconducting $T_{c}$. We assume the existence of a soft mode instability (ferroelectric-type transition) characterized by the typical Curie-Weiss behavior as a function of an external control parameter $n$:
\begin{equation}
\mathrm{Re}(\omega)=\sqrt{|n_c - n|}\,.
\label{Curie}
\end{equation}
In standard ferroelectrics, $n$ is of course the temperature $T$. For a derivation of the Curie-Weiss law with $n \equiv T$ induced by giant phonon anharmonicity (as in many thermoelectric materials~\cite{Delaire2015}) see Ref.~\cite{Casella}.
In ferroelectric-type SCs, $n$ could be electron doping~\cite{Ma2021}
or the mechanical strain\cite{Balatsky}, $n\equiv \epsilon$.
In both cases, the variation of the control parameter $n$ changes the phonon damping $\Gamma$, as mentioned above.

Using Eq.\eqref{Curie} together with Eq.\eqref{realpart}, we solve for the damping and obtain the following relationship:
\begin{equation}
\Gamma(n) = 2\sqrt{\omega_{0}^{2} - |n_c - n|}\,.
\label{damping}
\end{equation}

\begin{figure}
    \centering
    \includegraphics[width=0.75 \linewidth]{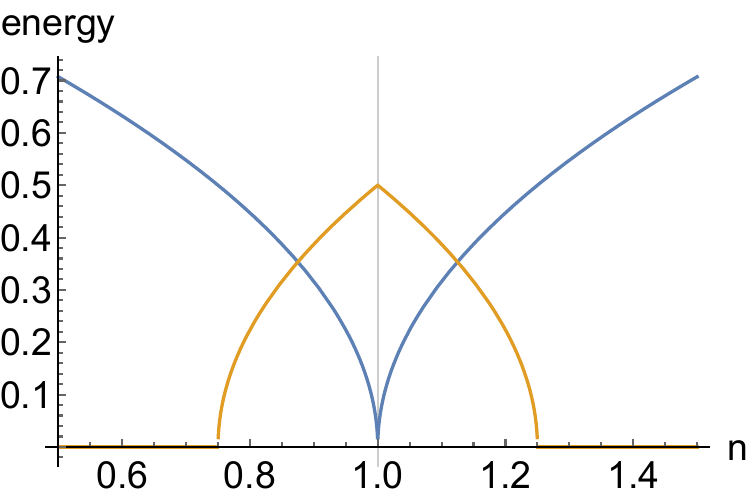}
    \caption{Simplified model expressions for the optical mode energy (\color{bluemathematica}\textbf{blue}\color{black}) and the inverse lifetime $\tau^{-1}$ (\color{orangemathematica}\textbf{orange}\color{black}) upon approaching the soft mode instability as a function of a generic control parameter $n$. Here, $\omega_0=0.5,n_c=1$.}
    \label{fig:model}
\end{figure}

The energy of the mode (real part of the frequency, Eq.\eqref{Curie}) and the inverse lifetime (imaginary part, Eq.\eqref{damping}) are plotted in Fig.\ref{fig:model} for the choice $\omega_0=0.5,n_c=1$.

It is important to stress that the Curie law in Eq.\eqref{Curie} is valid only close to the instability $n \sim n_c$ and it cannot be trusted far away from there. This is also evident from the expression for the damping in Eq.\eqref{damping} which stops to be real outside the window $|n_c-\omega_0^2|<n<n_c+\omega_0^2$. As a matter of fact, we will consider the external parameter $n$ only within that range in the rest of the manuscript.
\subsection{Gap equation with anharmonic phonon damping}
We briefly recall the anharmonic BCS framework of Ref.~\cite{Setty2021}. We assume a large electron density limit and isotropic pairing without explicitly taking into account strong coupling or multiband effects. The latter could be relevant to the specific case of SrTiO$_3$ and related superconductors. Such effects generally modify the superconducting gap, $T_c$ and chemical potential away from weak coupling values\cite{Chubukov2016}. In addition, we neglect the role of repulsive coulomb interactions in driving anisotropic pairing. The interplay of strong coupling effects with repulsive interactions and anisotropic pairing can engender novel superconducting phases that spontaneously break lattice symmetries-\cite{Setty2021-FPDW}. Multiple bands further modify the aforementioned properties\cite{Chubukov2016} and are expected to greatly enrich the physics we describe below. We leave generalizations to include these non-trivial effects for later work.   
 For a generic Fermionic Matsubara frequency $\omega_n$ and momentum $\textbf k$, we denote the superconducting gap function as $\Delta(i\omega_m,\textbf k)$. We assume throughout a quadratic nearly-free electron dispersion relation for the electronic band. With a constant coupling $g$, the gap equation can be derived from the Eliashberg equations in the one-loop (weak coupling) approximation, and takes the form~\cite{Marsiglio,Kleinert}
\begin{eqnarray}\nonumber
\Delta(i\omega_n, \textbf{k}) &=& \frac{g^2}{\beta V} \sum_{\textbf{q}, \omega_m} \frac{\Delta(i\omega_m, \textbf{k}+ \textbf{q}) \Pi(\textbf{q}, i\omega_n - i \omega_m)}{\omega_m^2 + \xi_{\textbf{k} + \textbf{q}}^2 + \Delta(i\omega_m, \textbf{k}+ \textbf{q})^2}\,,\\
&&
\label{Sum-GapEqn}
\end{eqnarray}
where $\beta$ is the inverse temperature and $V$ is the volume. In Matsubara frequency space, we choose the pairing mediator to be a damped optical phonon given by the bosonic propagator~\cite{Ziman}
\begin{equation}
\Pi(\textbf{q}, i \Omega_n) = \frac{1}{\Omega(q)^2 + \Omega_n^2 + \Gamma(q)\Omega_n}.
\label{propagator}
\end{equation}
Here $\Omega_n$ is the bosonic Matsubara frequency for the phonon, $\Omega(q) = \omega_0 $ is the frequency of the optical phonon. For our initial treatment, we neglect any dispersion effects of the optical phonon. We take the damping $\Gamma(q)$ equal to Eq.\eqref{damping} in order to account for the effect of the soft mode. As the phonon mode is optical, we choose no additional dispersion effects in the damping coefficient~\cite{Klemens1966}.

Assuming an isotropic, frequency-independent gap $\Delta(i\omega_n, \textbf k)\equiv\Delta$, we set the external frequency and momentum to zero without any loss of generality. Converting the resulting summation into an energy integral (and assuming a quadratic dispersion relation for the fermions), the gap equation becomes \\

\begin{eqnarray} \nonumber
1 &=& \sum_{m} \int_{-\infty}^{\infty}\frac{\lambda T d\xi}{\left[  \omega_0^2 + \Omega_m^2 - \Gamma\Omega_m\right] \left[\omega_m^2 + \xi^2 + \Delta^2\right]}\,,\\
&&
\label{Integral-GapEqn}
\end{eqnarray}
where the chemical potential is considered to be large so that the lower limit of the energy integral is taken to negative infinity. We also define the effective coupling constant $\lambda = N(0)g^2$ and  $N(0)$ is the density of states at the Fermi level. We can now perform the energy integral exactly in the limit of large chemical potential to obtain a simplified gap equation. The condition for $T_c$ can then be evaluated  by setting $\Delta=0$ to get
\begin{eqnarray}\nonumber
    1 &=&\sum_{m=-\infty}^{\infty} \frac{\bar{\lambda }}{2 \left|m+ \frac{1}{2} \right|\left(m^2 \bar{T}_c^2-m \bar{\Gamma } \bar{T}_c+1\right)}.\\
    &&
    \label{MSum}
\end{eqnarray}
Here $\bar{T}_c \equiv \frac{T_c}{\omega_0}$, $\bar{\Gamma} \equiv \frac{\Gamma}{\omega_0}$ and $\bar{\lambda} \equiv \frac{\lambda}{\omega_0}$ are the critical temperature, damping and the effective coupling constant normalized by the phonon frequency. Performing the Matsubara sum using methods described in Ref.~\cite{Varlamov2005} leads to an equation for $T_c$ that can be numerically solved. This condition is given as
\begin{widetext}
\begin{eqnarray}\nonumber
&&4 \Bigg(1-\frac{4 \gamma  \bar{\lambda }}{2 \bar{\Gamma } \bar{T}_c+\bar{T}_c^2+4}-\frac{4 \log (4) \bar{\lambda }}{2 \bar{\Gamma }
   \bar{T}_c+\bar{T}_c^2+4}-\frac{\bar{\lambda } \psi ^{(0)}\left(\frac{\bar{\Gamma }}{2 \bar{T}_c}-\frac{\sqrt{\bar{\Gamma }^2-4}}{2
   \bar{T}_c}+1\right)}{\bar{T}_c\sqrt{\bar{\Gamma }^2-4} -\bar{\Gamma }^2+\bar{\Gamma }\sqrt{\bar{\Gamma }^2-4} +4}-\frac{\bar{\lambda } \psi
   ^{(0)}\left(\frac{\sqrt{\bar{\Gamma }^2-4}}{2 \bar{T}_c}-\frac{\bar{\Gamma }}{2 \bar{T}_c}\right)}{\bar{T}_c\sqrt{\bar{\Gamma }^2-4} -\bar{\Gamma
   }^2+\bar{\Gamma }\sqrt{\bar{\Gamma }^2-4} +4}\\ 
   &&+\frac{\bar{\lambda } \psi ^{(0)}\left(-\frac{\bar{\Gamma }}{2 \bar{T}_c}-\frac{\sqrt{\bar{\Gamma }^2-4}}{2
   \bar{T}_c}\right)}{\bar{T}_c\sqrt{\bar{\Gamma }^2-4} +\bar{\Gamma }^2+\bar{\Gamma }\sqrt{\bar{\Gamma }^2-4} -4}+\frac{\bar{\lambda } \psi
   ^{(0)}\left(\frac{\bar{\Gamma }}{2 \bar{T}_c}+\frac{\sqrt{\bar{\Gamma }^2-4}}{2 \bar{T}_c}+1\right)}{\bar{T}_c\sqrt{\bar{\Gamma }^2-4} +\bar{\Gamma
   }^2+\bar{\Gamma }\sqrt{\bar{\Gamma }^2-4} -4}\Bigg) =0\,, \label{GapEqn}
   \end{eqnarray}
\end{widetext}
where $\psi^{(0)}\left(x\right)$ is the digamma function and $\gamma$ is the Euler constant. It should be noted that the Eq.\eqref{GapEqn} is always real (even when $\bar{\Gamma}^2-4<0$), due to the mutual cancellation of imaginary terms.
\begin{figure}[t]
    \centering
    \includegraphics[width=0.65\linewidth]{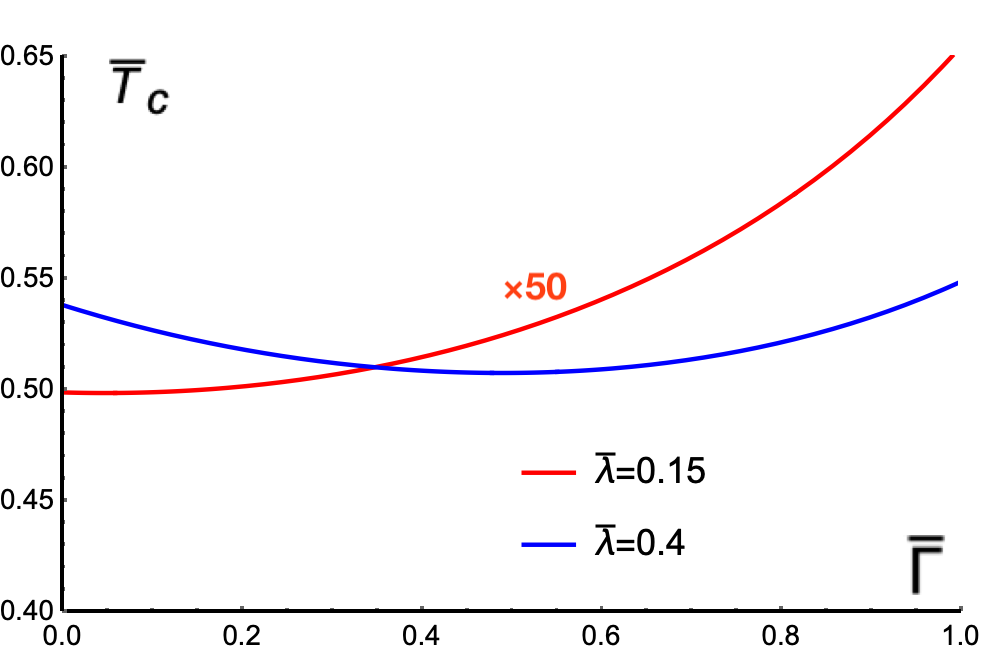}
    \caption{Plot of normalized critical temperature with damping $\Gamma$ for different effective couplings $\bar{\lambda}$.  For large values of $\bar{\Gamma}$, there is no solution for superconductivity.}
    \label{TcVsGamma}
\end{figure}

\begin{figure}[t]
    \centering
    \includegraphics[width=0.65\linewidth]{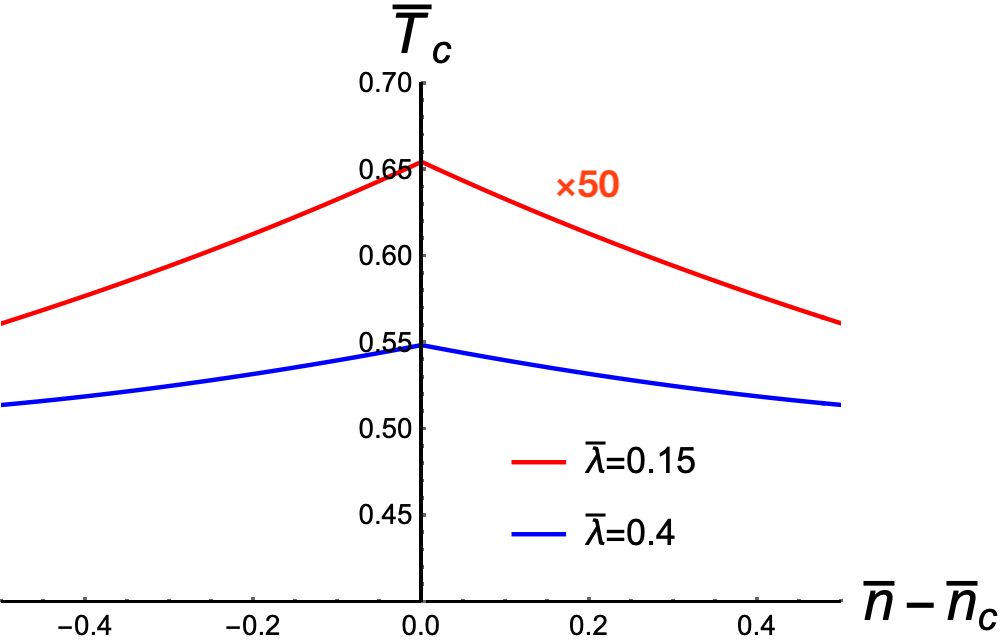}
    \caption{Plot of normalized critical temperature around the ferroelectric critical point $\bar n-\bar n_c$ for different effective couplings $\bar{\lambda}$. }
    \label{TcVsn}
\end{figure}
The solution of the gap equation, Eq.~\ref{GapEqn}, for $\bar T_c$ as a function of $\bar{\Gamma}$ is plotted in Fig.~\ref{TcVsGamma}. The curves exhibit two kinds of behaviors. The first is a monotonically increasing behavior of $\bar T_c$  for small $\bar {\lambda}$ (\color{red}\textbf{red} \color{black} curve). The second is a non-monotonic behavior of $\bar T_c$ with $\bar{\Gamma}$ for larger $\bar{\lambda}$ (\color{blue}\textbf{blue} \color{black} curve). In both cases, there is no solution for superconductivity for large enough $\bar{\Gamma}\gtrapprox 1$, although the exact critical value is different for various $\bar{\lambda}$. The behavior of the $\bar{T}_c$ curves for different $\bar{\lambda}$ can be understood in several ways. First, we note that there are two types of Matsubara frequency transfers that play a role in the gap Eq.~\ref{MSum}: positive ($m>0$, Stokes) and  negative ($m<0$, anti-Stokes) contributions.  The scattering amplitudes between the electrons and the Stokes (anti-Stokes) phonons is enhanced (reduced) by damping effects from the denominator of the gap equation. Hence, for a given coupling $\bar{\lambda}$ and damping $\bar{\Gamma}$, Stokes (anti-Stokes) phonons lead to a larger (smaller) pair binding between electrons and require a larger (smaller) critical temperature to satisfy the gap equation. For small $\bar{\lambda}$, the Stokes scattering dominates for all $\bar{\Gamma}$ and a monotonically increasing $\bar T_c$ is obtained. But for larger $\bar{\lambda}$, there is a competition between the Stokes and anti-Stokes scatterings leading to a minimum in $\bar T_c$ at non-zero $\bar{\Gamma}$. The behavior of the $\bar T_c$ curves can also be understood from a perturbative expansion of the gap equation for small $\bar{\Gamma}$ up to second order in $\bar{\Gamma}$. The linear order term $O(\bar{\Gamma})$ is sensitive to electron scattering from Stokes and anti-Stokes phonons with the latter contributions dominating to cause a decrease in $\bar{T}_c$. On the other hand, the quadratic $O(\bar{\Gamma}^2)$ term is always positive and mostly insensitive to the sign of the phonon energy exchange acting to enhance pair binding and $\bar T_c$. Below a critical $\bar{\lambda}$, the $O(\bar{\Gamma}^2)$ dominates over $O(\bar{\Gamma})$ for all $\bar{\Gamma}$ giving a monotonic increase in $\bar{T}_c$ (this is due to the $\bar{\lambda}$-dependent prefactors in the $\Gamma$-expansion), while for larger $\bar{\lambda}$ the two orders compete resulting in a minimum of $\bar T_c$ at a finite $\bar{\Gamma}$. The lack of superconductivity for large $\bar{\Gamma}$ can be understood by taking the limit $\bar{\Gamma} \gg 1$ in Eq.~\ref{MSum}. In this limit, the leading order contribution to the gap equation diverges for $m=0$ and hence there is no solution for $T_c$ possible.   
\section{Results}
\subsection{Emergence of the superconducting dome}
\begin{figure}[t]
    \centering
     \includegraphics[width=0.45\linewidth]{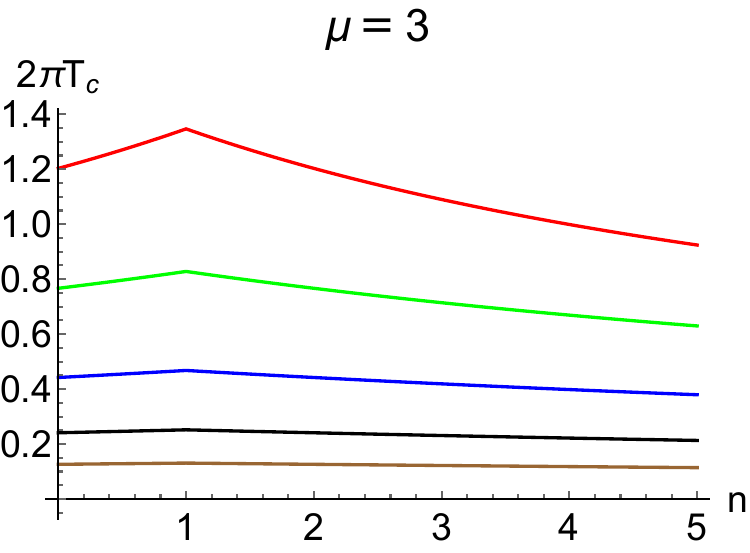}\quad
      \includegraphics[width=0.45\linewidth]{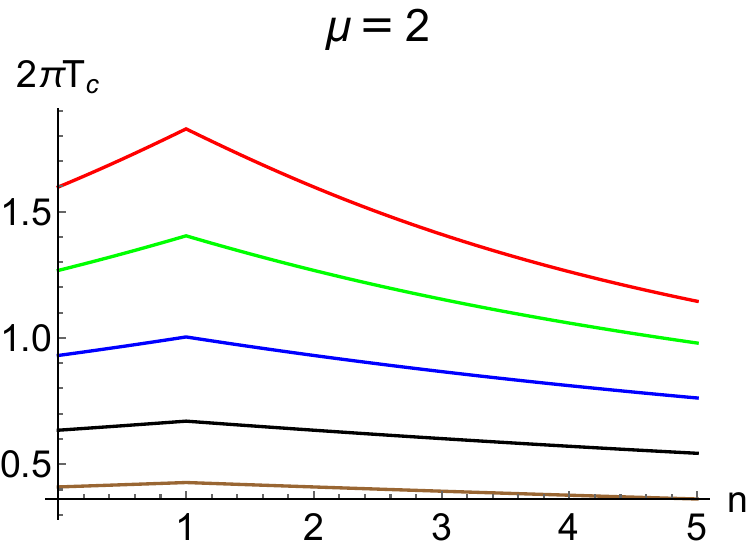}\quad
      
           \vspace{0.3cm}
           
      \includegraphics[width=0.45\linewidth]{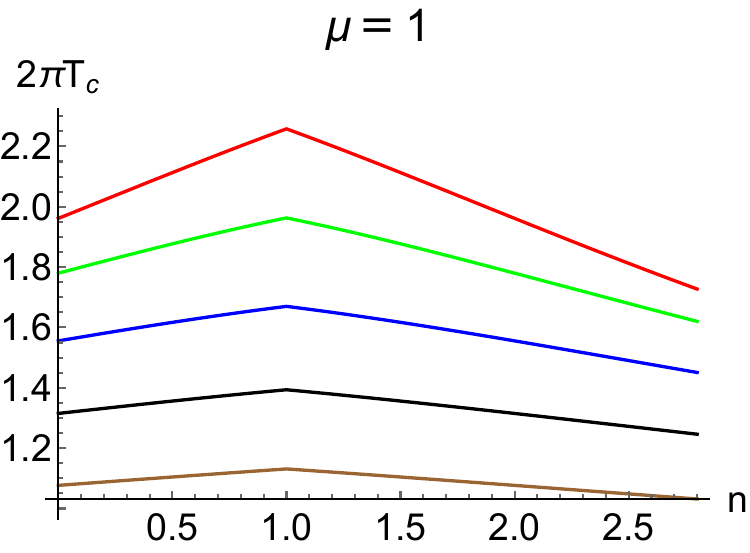}\quad
      \includegraphics[width=0.45\linewidth]{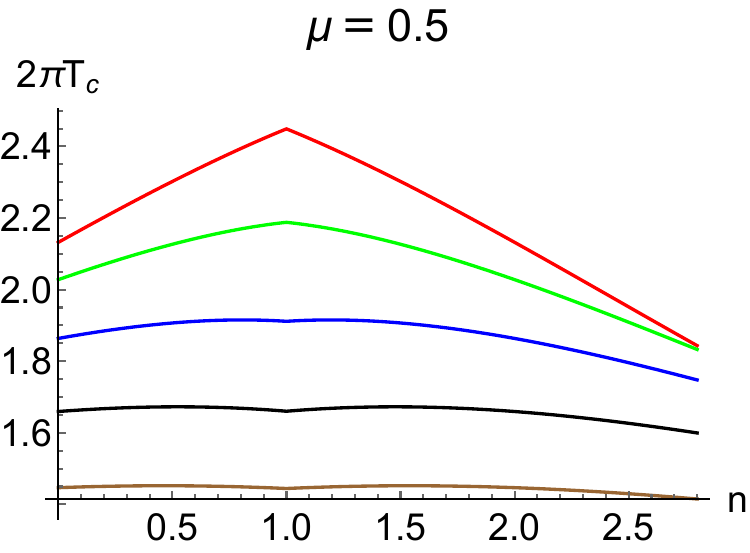}
    \caption{Superconducting dome predicted by the anharmonic theoretical model with a dispersion of the soft phonon quadratic in momentum transfer, i.e. $\Omega(q)=\omega_{0} + \alpha q^{2}$, where coefficient $\alpha$ is the stiffness or ``velocity''. $T_c$ is plotted as a function of the soft phonon control parameter $n$ upon approaching the soft mode instability at $n=n_c =1$. In each plot, curves are plotted for the following values of the optical energy $\omega_{0}$, from top to bottom: $\omega_0 = 0.3,0.4, 0.5, 0.6, 0.7$. In each plot, the chemical potential $\mu$ is fixed to the value shown.}
    \label{Stiffness}
\end{figure}

Upon using Eq.~\eqref{damping} for the damping coefficient $\Gamma(n)$ in Eq.~\eqref{GapEqn} and solving numerically, we can study the evolution of $T_c$ as a function of the external control parameter $n$ that drives the ferroelectric transition. The trend is shown in Fig.~\ref{TcVsn}.
Clearly the theoretical calculations display a superconducting dome centered on the soft mode at $n=n_{c}$.
In particular, the $T_c$ grows with approaching the soft mode instability towards $n=n_c$ on both sides, i.e. both above and below $n_c$; or in other words, both in the ferroelectric and in the paraelectric phase.
A physical explanation for this widely observed behavior~\cite{Spaldin,Balatsky,Ma2021} is thus offered by the structure of our model.

The mechanism giving rise to the $T_c$ trend in Fig.~\ref{TcVsn} can be understood in the following manner. By considering Fig.\ref{fig:model}, it is clear that the damping $\Gamma$ grows with approaching $n=n_c$ on both sides of the transition, in qualitative agreement with experiments on ferroelectrics~\cite{Delaire2020}.
 As $n$ approaches $n_c$ in Eq.~\ref{damping}, $\bar{\Gamma}$ becomes larger and approaches a value of $O(1)$. In the vicinity of this limit, $\bar T_c$ increases with increasing $\bar{\Gamma}$ regardless of the coupling $\bar{\lambda}$ (see Fig.~\ref{TcVsGamma}). Hence, this correlation of $T_c$ with $\bar{\Gamma}$ near the critical point leads to a reduction of $T_c$ away from $n_c$ when $\bar{\Gamma}$ decreases.  The resulting dome-like behavior of $T_c$ vs $n$ is shown in Fig.~\ref{TcVsn}.  The dome like feature is therefore ultimately tied to the dominant, damping-assisted, Stokes scattering processes which act as to increase $T_c$, by overcoming the de-pairing effects from anti-Stokes processes. This regime also coincides with a value of $\bar{\Gamma}$ that is small enough to perform a perturbative series expansion in the parameter $\bar{\Gamma}<1$, but also large enough so that second order ($O(\bar{\Gamma}^2$)) damping processes dominate over the linear order terms ($O(\bar{\Gamma})$) in the gap equation. Notice that this is possible only because the coefficients of the linear and quadratic terms in the perturbative expansion $T_c(\bar{\Gamma})=T_c(0)+ a \bar{\Gamma} + b \bar{\Gamma}^2+O(\bar{\Gamma}^3)$ are non-trivial functions of the various parameters (e.g. $\bar{\lambda}$). Their ratio $a/b$ can be arbitrarily smaller than $1$ allowing for a window $a/b<\bar{\Gamma}<1$ in which the quadratic contribution is leading as mentioned above. \color{black} {The second order processes lead to a $T_c$ enhancement as described in the previous section. \par
 The physics of ferroelectric superconductors described above must be contrasted with previously studied anharmonic~\cite{Setty2020, Setty2021} or glassy~\cite{Setty2019} systems where the pairing mediator contains a non-trivial spatial stiffness term associated with some dispersive $q$-dependence of the phonon. First, due to the key role of acoustic phonons in those studies, a non-trivial spatial stiffness is important to achieve enhanced $T_c$ with damping or decoherence. In the current scenario where the damping coefficient $\bar{\Gamma}$ is independent of momentum transfer due to the nature of the optical mode, a non-trivial spatial stiffness is not necessary to achieve $T_c$ enhancement with $\bar{\Gamma}$. Second, the mechanism of $T_c$ enhancement in damped acoustic or glassy systems involves a constructive interference either between low and high energy phonons or Stokes--anti-Stokes processes. In the current theory we propose for ferroelectric materials, the $T_c$ enhancement occurs mainly from dominant Stokes processes while anti-Stokes processes are detrimental to superconductivity.  Finally, the central role of a non-trivial spatial stiffness in Cooper pairing mediated by acoustic, optical or glassy modes is to enhance superconductivity for perturbatively \textit{weak} anharmonic decoherence only (even for intermediate $\bar{\lambda}$). While $T_c$ peaks at an optimal damping scale, set by the spatial stiffness of the bosonic mode, it is gradually suppressed for very strong damping. Consequently, the anticorrelation between strong damping and superconductivity  gives a dip in $T_c$ at $n_c$ rather than a dome,  not quite consistent with experiments in ferroelectric materials. We therefore do not anticipate a central role for a pairing propagator with finite spatial stiffness (due to non-trivial phonon dispersion in momentum space), for understanding superconductivity in ferroelectric materials near a ferroelectric critical point. In the current proposal, however, $T_c$ is enhanced for strong enough damping with an abrupt loss of superconductivity above a critical damping set by $\bar{\lambda}$, thus yielding the necessary experimental dependence on $n$. Therefore, material design principles that aim to realize this effect experimentally in other realistic superconducting ferroelectric systems can be guided by regions of the phase diagram with significantly damped bosonic modes. A weak coupling constant (well within the BCS limit), in addition, ensures that $T_c$ monotonically increases with damping up to a critical value, hence ensuring a peak at the critical point.

Before we conclude, for the sake of completeness we briefly discuss the possibility of obtaining the experimental $T_c$ curves near $n_c$ in the weak dissipation limit and intermediate $\bar{\lambda}$. This requires a $T_c$ enhancement with dissipation at $\bar{\lambda}\sim 0.5$. Under this working assumption, we must include a dispersion term of the optical phonon which is quadratic in the momentum transfer $q$ with a prefactor coefficient $\alpha$ that sets the stiffness or ``velocity''~\cite{Setty2021}, i.e. $\Omega(q) = \omega_0 + \alpha \,q^{2}$.  Numerically evaluating the gap equation as a function of $n-n_c$, we plot the resulting $T_c$ behavior in Fig.~\ref{Stiffness}. An explicit dependence on the chemical potential $\mu$ appears due to the quadratic dispersion term of the optical phonon. While the role of $\mu$ is to reduce the overall scale of $T_c$ (the chemical potential acts like a mass term and hence reduces $T_c$), the general dome-like trend of $T_c$ with $n$ is preserved for small enough $\omega_0$. 
The experimental optical phonon frequencies in SrTiO$_{3}$ go as low as a few meV (40 cm$^{-1}$) and hence well within the regime of our interest (cfr. \cite{Himmetoglu}). Despite the small Fermi energy in SrTiO$_{3}$, we expect the smallness of $\omega_0$ to  be well within the dome-like region. We must note that for small $\mu$, the current theory will have to be modified to include strong coupling effects.
For large $\omega_0$ and small $\mu$, the dome gives way to a shallow dip in $T_c$ at the critical point.

\vspace*{0.2cm} 

\section{Conclusion}
We presented a theory of superconductivity enhanced by anharmonic damping in ferroelectric-type materials based on a minimal agnostic model for the soft mode instability.
The key effect driving the enhancement of $T_c$ at the soft mode transition is the anharmonic phonon damping which increases upon approaching the transition on both sides, hence the ubiquitous dome. The dome-like feature stems from the dominant, damping-assisted, Stokes scattering processes which act as to increase $T_c$, by overcoming the de-pairing effects from anti-Stokes processes. Treating the damping effects perturbatively, we showed that the quadratic order damping processes are key to enhancing the pairing as opposed to linear order terms. 
This explains the crucial role of phonon anharmonic damping near the ferroelectric soft mode to obtain the experimentally observed behavior of $T_c$, which is supported by gigantic values of anharmonic Gr\"{u}neisen parameter observed in these systems\cite{Balatsky}. The presented theory explains the widely observed dome in $T_c$ observed at ferroelectric-type transitions purely in terms of lattice dynamics and phonon physics, and without the need of invoking any exotic electronic effects. Furthermore, it provides useful guidelines for material-by-design rules to engineer ferroelectric-type materials with optimized superconducting properties.

\textit{Acknowledgements} -- M.B. acknowledges the support of the  Shanghai Municipal Science and Technology Major Project (Grant No.2019SHZDZX01). C.S. is supported by the U.S. DOE grant number DE-FG02-05ER46236. A.Z. acknowledges financial support from US Army Research Laboratory and US Army Research Office through contract nr. W911NF-19-2-0055.

\bibliography{anharmonicity2}

\begin{thebibliography}{61}%
\makeatletter
\providecommand \@ifxundefined [1]{%
 \@ifx{#1\undefined}
}%
\providecommand \@ifnum [1]{%
 \ifnum #1\expandafter \@firstoftwo
 \else \expandafter \@secondoftwo
 \fi
}%
\providecommand \@ifx [1]{%
 \ifx #1\expandafter \@firstoftwo
 \else \expandafter \@secondoftwo
 \fi
}%
\providecommand \natexlab [1]{#1}%
\providecommand \enquote  [1]{``#1''}%
\providecommand \bibnamefont  [1]{#1}%
\providecommand \bibfnamefont [1]{#1}%
\providecommand \citenamefont [1]{#1}%
\providecommand \href@noop [0]{\@secondoftwo}%
\providecommand \href [0]{\begingroup \@sanitize@url \@href}%
\providecommand \@href[1]{\@@startlink{#1}\@@href}%
\providecommand \@@href[1]{\endgroup#1\@@endlink}%
\providecommand \@sanitize@url [0]{\catcode `\\12\catcode `\$12\catcode
  `\&12\catcode `\#12\catcode `\^12\catcode `\_12\catcode `\%12\relax}%
\providecommand \@@startlink[1]{}%
\providecommand \@@endlink[0]{}%
\providecommand \url  [0]{\begingroup\@sanitize@url \@url }%
\providecommand \@url [1]{\endgroup\@href {#1}{\urlprefix }}%
\providecommand \urlprefix  [0]{URL }%
\providecommand \Eprint [0]{\href }%
\providecommand \doibase [0]{http://dx.doi.org/}%
\providecommand \selectlanguage [0]{\@gobble}%
\providecommand \bibinfo  [0]{\@secondoftwo}%
\providecommand \bibfield  [0]{\@secondoftwo}%
\providecommand \translation [1]{[#1]}%
\providecommand \BibitemOpen [0]{}%
\providecommand \bibitemStop [0]{}%
\providecommand \bibitemNoStop [0]{.\EOS\space}%
\providecommand \EOS [0]{\spacefactor3000\relax}%
\providecommand \BibitemShut  [1]{\csname bibitem#1\endcsname}%
\let\auto@bib@innerbib\@empty
\bibitem [{\citenamefont {Matthias}(1955)}]{Matthias1955}%
  \BibitemOpen
  \bibfield  {author} {\bibinfo {author} {\bibfnamefont {B.~T.}\ \bibnamefont
  {Matthias}},\ }\href@noop {} {\bibfield  {journal} {\bibinfo  {journal}
  {Physical review}\ }\textbf {\bibinfo {volume} {97}},\ \bibinfo {pages} {74}
  (\bibinfo {year} {1955})}\BibitemShut {NoStop}%
\bibitem [{\citenamefont {Anderson}\ and\ \citenamefont
  {Blount}(1965)}]{Blount1965}%
  \BibitemOpen
  \bibfield  {author} {\bibinfo {author} {\bibfnamefont {P.~W.}\ \bibnamefont
  {Anderson}}\ and\ \bibinfo {author} {\bibfnamefont {E.}~\bibnamefont
  {Blount}},\ }\href@noop {} {\bibfield  {journal} {\bibinfo  {journal}
  {Physical Review Letters}\ }\textbf {\bibinfo {volume} {14}},\ \bibinfo
  {pages} {217} (\bibinfo {year} {1965})}\BibitemShut {NoStop}%
\bibitem [{\citenamefont {Appel}(1966)}]{Appel1966}%
  \BibitemOpen
  \bibfield  {author} {\bibinfo {author} {\bibfnamefont {J.}~\bibnamefont
  {Appel}},\ }\href@noop {} {\bibfield  {journal} {\bibinfo  {journal}
  {Physical Review Letters}\ }\textbf {\bibinfo {volume} {17}},\ \bibinfo
  {pages} {1045} (\bibinfo {year} {1966})}\BibitemShut {NoStop}%
\bibitem [{\citenamefont {Appel}(1969)}]{Appel1969}%
  \BibitemOpen
  \bibfield  {author} {\bibinfo {author} {\bibfnamefont {J.}~\bibnamefont
  {Appel}},\ }\href@noop {} {\bibfield  {journal} {\bibinfo  {journal}
  {Physical Review}\ }\textbf {\bibinfo {volume} {180}},\ \bibinfo {pages}
  {508} (\bibinfo {year} {1969})}\BibitemShut {NoStop}%
\bibitem [{\citenamefont {Ginzburg}(1987)}]{Ginzburg1987}%
  \BibitemOpen
  \bibfield  {author} {\bibinfo {author} {\bibfnamefont {V.~L.}\ \bibnamefont
  {Ginzburg}},\ }\href {\doibase 10.1080/00150198708009019} {\bibfield
  {journal} {\bibinfo  {journal} {Ferroelectrics}\ }\textbf {\bibinfo {volume}
  {76}},\ \bibinfo {pages} {3} (\bibinfo {year} {1987})},\ \Eprint
  {http://arxiv.org/abs/https://doi.org/10.1080/00150198708009019}
  {https://doi.org/10.1080/00150198708009019} \BibitemShut {NoStop}%
\bibitem [{\citenamefont {Bussmann-Holder}\ \emph {et~al.}(1989)\citenamefont
  {Bussmann-Holder}, \citenamefont {Simon},\ and\ \citenamefont
  {B{\"u}ttner}}]{Buttner1989}%
  \BibitemOpen
  \bibfield  {author} {\bibinfo {author} {\bibfnamefont {A.}~\bibnamefont
  {Bussmann-Holder}}, \bibinfo {author} {\bibfnamefont {A.}~\bibnamefont
  {Simon}}, \ and\ \bibinfo {author} {\bibfnamefont {H.}~\bibnamefont
  {B{\"u}ttner}},\ }\href@noop {} {\bibfield  {journal} {\bibinfo  {journal}
  {Physical Review B}\ }\textbf {\bibinfo {volume} {39}},\ \bibinfo {pages}
  {207} (\bibinfo {year} {1989})}\BibitemShut {NoStop}%
\bibitem [{\citenamefont {Binnig}\ \emph {et~al.}(1980)\citenamefont {Binnig},
  \citenamefont {Baratoff}, \citenamefont {Hoenig},\ and\ \citenamefont
  {Bednorz}}]{Binnig}%
  \BibitemOpen
  \bibfield  {author} {\bibinfo {author} {\bibfnamefont {G.}~\bibnamefont
  {Binnig}}, \bibinfo {author} {\bibfnamefont {A.}~\bibnamefont {Baratoff}},
  \bibinfo {author} {\bibfnamefont {H.~E.}\ \bibnamefont {Hoenig}}, \ and\
  \bibinfo {author} {\bibfnamefont {J.~G.}\ \bibnamefont {Bednorz}},\ }\href
  {\doibase 10.1103/PhysRevLett.45.1352} {\bibfield  {journal} {\bibinfo
  {journal} {Phys. Rev. Lett.}\ }\textbf {\bibinfo {volume} {45}},\ \bibinfo
  {pages} {1352} (\bibinfo {year} {1980})}\BibitemShut {NoStop}%
\bibitem [{\citenamefont {Enderlein}\ \emph {et~al.}(2020)\citenamefont
  {Enderlein}, \citenamefont {de~Oliveira}, \citenamefont {Tompsett},
  \citenamefont {Saitovitch}, \citenamefont {Saxena}, \citenamefont
  {Lonzarich},\ and\ \citenamefont {Rowley}}]{Lonzarich2020}%
  \BibitemOpen
  \bibfield  {author} {\bibinfo {author} {\bibfnamefont {C.}~\bibnamefont
  {Enderlein}}, \bibinfo {author} {\bibfnamefont {J.~F.}\ \bibnamefont
  {de~Oliveira}}, \bibinfo {author} {\bibfnamefont {D.~A.}\ \bibnamefont
  {Tompsett}}, \bibinfo {author} {\bibfnamefont {E.~B.}\ \bibnamefont
  {Saitovitch}}, \bibinfo {author} {\bibfnamefont {S.~S.}\ \bibnamefont
  {Saxena}}, \bibinfo {author} {\bibfnamefont {G.~G.}\ \bibnamefont
  {Lonzarich}}, \ and\ \bibinfo {author} {\bibfnamefont {S.~E.}\ \bibnamefont
  {Rowley}},\ }\href {\doibase 10.1038/s41467-020-18438-0} {\bibfield
  {journal} {\bibinfo  {journal} {Nature Communications}\ }\textbf {\bibinfo
  {volume} {11}},\ \bibinfo {pages} {4852} (\bibinfo {year}
  {2020})}\BibitemShut {NoStop}%
\bibitem [{\citenamefont {Koonce}\ \emph {et~al.}(1967)\citenamefont {Koonce},
  \citenamefont {Cohen}, \citenamefont {Schooley}, \citenamefont {Hosler},\
  and\ \citenamefont {Pfeiffer}}]{Pfeiffer1967}%
  \BibitemOpen
  \bibfield  {author} {\bibinfo {author} {\bibfnamefont {C.}~\bibnamefont
  {Koonce}}, \bibinfo {author} {\bibfnamefont {M.~L.}\ \bibnamefont {Cohen}},
  \bibinfo {author} {\bibfnamefont {J.}~\bibnamefont {Schooley}}, \bibinfo
  {author} {\bibfnamefont {W.}~\bibnamefont {Hosler}}, \ and\ \bibinfo {author}
  {\bibfnamefont {E.}~\bibnamefont {Pfeiffer}},\ }\href@noop {} {\bibfield
  {journal} {\bibinfo  {journal} {Physical Review}\ }\textbf {\bibinfo {volume}
  {163}},\ \bibinfo {pages} {380} (\bibinfo {year} {1967})}\BibitemShut
  {NoStop}%
\bibitem [{\citenamefont {Edge}\ \emph {et~al.}(2015)\citenamefont {Edge},
  \citenamefont {Kedem}, \citenamefont {Aschauer}, \citenamefont {Spaldin},\
  and\ \citenamefont {Balatsky}}]{Spaldin}%
  \BibitemOpen
  \bibfield  {author} {\bibinfo {author} {\bibfnamefont {J.~M.}\ \bibnamefont
  {Edge}}, \bibinfo {author} {\bibfnamefont {Y.}~\bibnamefont {Kedem}},
  \bibinfo {author} {\bibfnamefont {U.}~\bibnamefont {Aschauer}}, \bibinfo
  {author} {\bibfnamefont {N.~A.}\ \bibnamefont {Spaldin}}, \ and\ \bibinfo
  {author} {\bibfnamefont {A.~V.}\ \bibnamefont {Balatsky}},\ }\href {\doibase
  10.1103/PhysRevLett.115.247002} {\bibfield  {journal} {\bibinfo  {journal}
  {Phys. Rev. Lett.}\ }\textbf {\bibinfo {volume} {115}},\ \bibinfo {pages}
  {247002} (\bibinfo {year} {2015})}\BibitemShut {NoStop}%
\bibitem [{\citenamefont {Schooley}\ \emph {et~al.}(1964)\citenamefont
  {Schooley}, \citenamefont {Hosler},\ and\ \citenamefont {Cohen}}]{Cohen1964}%
  \BibitemOpen
  \bibfield  {author} {\bibinfo {author} {\bibfnamefont {J.~F.}\ \bibnamefont
  {Schooley}}, \bibinfo {author} {\bibfnamefont {W.~R.}\ \bibnamefont
  {Hosler}}, \ and\ \bibinfo {author} {\bibfnamefont {M.~L.}\ \bibnamefont
  {Cohen}},\ }\href {\doibase 10.1103/PhysRevLett.12.474} {\bibfield  {journal}
  {\bibinfo  {journal} {Phys. Rev. Lett.}\ }\textbf {\bibinfo {volume} {12}},\
  \bibinfo {pages} {474} (\bibinfo {year} {1964})}\BibitemShut {NoStop}%
\bibitem [{\citenamefont {Lin}\ \emph {et~al.}(2013)\citenamefont {Lin},
  \citenamefont {Zhu}, \citenamefont {Fauqu\'e},\ and\ \citenamefont
  {Behnia}}]{Behnia2013}%
  \BibitemOpen
  \bibfield  {author} {\bibinfo {author} {\bibfnamefont {X.}~\bibnamefont
  {Lin}}, \bibinfo {author} {\bibfnamefont {Z.}~\bibnamefont {Zhu}}, \bibinfo
  {author} {\bibfnamefont {B.}~\bibnamefont {Fauqu\'e}}, \ and\ \bibinfo
  {author} {\bibfnamefont {K.}~\bibnamefont {Behnia}},\ }\href {\doibase
  10.1103/PhysRevX.3.021002} {\bibfield  {journal} {\bibinfo  {journal} {Phys.
  Rev. X}\ }\textbf {\bibinfo {volume} {3}},\ \bibinfo {pages} {021002}
  (\bibinfo {year} {2013})}\BibitemShut {NoStop}%
\bibitem [{\citenamefont {Chubukov}\ \emph {et~al.}(2016)\citenamefont
  {Chubukov}, \citenamefont {Eremin},\ and\ \citenamefont
  {Efremov}}]{Chubukov2016}%
  \BibitemOpen
  \bibfield  {author} {\bibinfo {author} {\bibfnamefont {A.~V.}\ \bibnamefont
  {Chubukov}}, \bibinfo {author} {\bibfnamefont {I.}~\bibnamefont {Eremin}}, \
  and\ \bibinfo {author} {\bibfnamefont {D.~V.}\ \bibnamefont {Efremov}},\
  }\href@noop {} {\bibfield  {journal} {\bibinfo  {journal} {Physical Review
  B}\ }\textbf {\bibinfo {volume} {93}},\ \bibinfo {pages} {174516} (\bibinfo
  {year} {2016})}\BibitemShut {NoStop}%
\bibitem [{\citenamefont {Ma}\ \emph {et~al.}(2021)\citenamefont {Ma},
  \citenamefont {Yang},\ and\ \citenamefont {Chen}}]{Ma2021}%
  \BibitemOpen
  \bibfield  {author} {\bibinfo {author} {\bibfnamefont {J.}~\bibnamefont
  {Ma}}, \bibinfo {author} {\bibfnamefont {R.}~\bibnamefont {Yang}}, \ and\
  \bibinfo {author} {\bibfnamefont {H.}~\bibnamefont {Chen}},\ }\href {\doibase
  10.1038/s41467-021-22541-1} {\bibfield  {journal} {\bibinfo  {journal}
  {Nature Communications}\ }\textbf {\bibinfo {volume} {12}},\ \bibinfo {pages}
  {2314} (\bibinfo {year} {2021})}\BibitemShut {NoStop}%
\bibitem [{\citenamefont {Franklin}\ \emph {et~al.}(2021)\citenamefont
  {Franklin}, \citenamefont {Xu}, \citenamefont {Davino}, \citenamefont
  {Mahabir}, \citenamefont {Balatsky}, \citenamefont {Aschauer},\ and\
  \citenamefont {Sochnikov}}]{Balatsky}%
  \BibitemOpen
  \bibfield  {author} {\bibinfo {author} {\bibfnamefont {J.}~\bibnamefont
  {Franklin}}, \bibinfo {author} {\bibfnamefont {B.}~\bibnamefont {Xu}},
  \bibinfo {author} {\bibfnamefont {D.}~\bibnamefont {Davino}}, \bibinfo
  {author} {\bibfnamefont {A.}~\bibnamefont {Mahabir}}, \bibinfo {author}
  {\bibfnamefont {A.~V.}\ \bibnamefont {Balatsky}}, \bibinfo {author}
  {\bibfnamefont {U.}~\bibnamefont {Aschauer}}, \ and\ \bibinfo {author}
  {\bibfnamefont {I.}~\bibnamefont {Sochnikov}},\ }\href {\doibase
  10.1103/PhysRevB.103.214511} {\bibfield  {journal} {\bibinfo  {journal}
  {Phys. Rev. B}\ }\textbf {\bibinfo {volume} {103}},\ \bibinfo {pages}
  {214511} (\bibinfo {year} {2021})}\BibitemShut {NoStop}%
\bibitem [{\citenamefont {Hameed}\ \emph {et~al.}(2021)\citenamefont {Hameed},
  \citenamefont {Pelc}, \citenamefont {Anderson}, \citenamefont {Klein},
  \citenamefont {Spieker}, \citenamefont {Yue}, \citenamefont {Das},
  \citenamefont {Ramberger}, \citenamefont {Lukas}, \citenamefont {Liu},
  \citenamefont {Krogstad}, \citenamefont {Osborn}, \citenamefont {Li},
  \citenamefont {Leighton}, \citenamefont {Fernandes},\ and\ \citenamefont
  {Greven}}]{Hameed2021}%
  \BibitemOpen
  \bibfield  {author} {\bibinfo {author} {\bibfnamefont {S.}~\bibnamefont
  {Hameed}}, \bibinfo {author} {\bibfnamefont {D.}~\bibnamefont {Pelc}},
  \bibinfo {author} {\bibfnamefont {Z.~W.}\ \bibnamefont {Anderson}}, \bibinfo
  {author} {\bibfnamefont {A.}~\bibnamefont {Klein}}, \bibinfo {author}
  {\bibfnamefont {R.~J.}\ \bibnamefont {Spieker}}, \bibinfo {author}
  {\bibfnamefont {L.}~\bibnamefont {Yue}}, \bibinfo {author} {\bibfnamefont
  {B.}~\bibnamefont {Das}}, \bibinfo {author} {\bibfnamefont {J.}~\bibnamefont
  {Ramberger}}, \bibinfo {author} {\bibfnamefont {M.}~\bibnamefont {Lukas}},
  \bibinfo {author} {\bibfnamefont {Y.}~\bibnamefont {Liu}}, \bibinfo {author}
  {\bibfnamefont {M.~J.}\ \bibnamefont {Krogstad}}, \bibinfo {author}
  {\bibfnamefont {R.}~\bibnamefont {Osborn}}, \bibinfo {author} {\bibfnamefont
  {Y.}~\bibnamefont {Li}}, \bibinfo {author} {\bibfnamefont {C.}~\bibnamefont
  {Leighton}}, \bibinfo {author} {\bibfnamefont {R.~M.}\ \bibnamefont
  {Fernandes}}, \ and\ \bibinfo {author} {\bibfnamefont {M.}~\bibnamefont
  {Greven}},\ }\href {\doibase 10.1038/s41563-021-01102-3} {\bibfield
  {journal} {\bibinfo  {journal} {Nature Materials}\ } (\bibinfo {year}
  {2021}),\ 10.1038/s41563-021-01102-3}\BibitemShut {NoStop}%
\bibitem [{\citenamefont {Gastiasoro}\ \emph
  {et~al.}(2020{\natexlab{a}})\citenamefont {Gastiasoro}, \citenamefont
  {Ruhman},\ and\ \citenamefont {Fernandes}}]{Gastiasoro}%
  \BibitemOpen
  \bibfield  {author} {\bibinfo {author} {\bibfnamefont {M.~N.}\ \bibnamefont
  {Gastiasoro}}, \bibinfo {author} {\bibfnamefont {J.}~\bibnamefont {Ruhman}},
  \ and\ \bibinfo {author} {\bibfnamefont {R.~M.}\ \bibnamefont {Fernandes}},\
  }\href {\doibase https://doi.org/10.1016/j.aop.2020.168107} {\bibfield
  {journal} {\bibinfo  {journal} {Annals of Physics}\ }\textbf {\bibinfo
  {volume} {417}},\ \bibinfo {pages} {168107} (\bibinfo {year}
  {2020}{\natexlab{a}})}\BibitemShut {NoStop}%
\bibitem [{\citenamefont {Ruhman}\ and\ \citenamefont {Lee}(2016)}]{Ruhman}%
  \BibitemOpen
  \bibfield  {author} {\bibinfo {author} {\bibfnamefont {J.}~\bibnamefont
  {Ruhman}}\ and\ \bibinfo {author} {\bibfnamefont {P.~A.}\ \bibnamefont
  {Lee}},\ }\href {\doibase 10.1103/PhysRevB.94.224515} {\bibfield  {journal}
  {\bibinfo  {journal} {Phys. Rev. B}\ }\textbf {\bibinfo {volume} {94}},\
  \bibinfo {pages} {224515} (\bibinfo {year} {2016})}\BibitemShut {NoStop}%
\bibitem [{\citenamefont {{Yu}}\ \emph {et~al.}(2021)\citenamefont {{Yu}},
  \citenamefont {{Hwang}}, \citenamefont {{Raghu}},\ and\ \citenamefont
  {{Chung}}}]{Koreans}%
  \BibitemOpen
  \bibfield  {author} {\bibinfo {author} {\bibfnamefont {Y.}~\bibnamefont
  {{Yu}}}, \bibinfo {author} {\bibfnamefont {H.~Y.}\ \bibnamefont {{Hwang}}},
  \bibinfo {author} {\bibfnamefont {S.}~\bibnamefont {{Raghu}}}, \ and\
  \bibinfo {author} {\bibfnamefont {S.~B.}\ \bibnamefont {{Chung}}},\
  }\href@noop {} {\bibfield  {journal} {\bibinfo  {journal} {arXiv e-prints}\
  ,\ \bibinfo {eid} {arXiv:2110.03710}} (\bibinfo {year} {2021})},\ \Eprint
  {http://arxiv.org/abs/2110.03710} {arXiv:2110.03710 [cond-mat.supr-con]}
  \BibitemShut {NoStop}%
\bibitem [{\citenamefont {Rosenstein}\ and\ \citenamefont
  {Shapiro}(2019)}]{Shapiro2019}%
  \BibitemOpen
  \bibfield  {author} {\bibinfo {author} {\bibfnamefont {B.}~\bibnamefont
  {Rosenstein}}\ and\ \bibinfo {author} {\bibfnamefont {B.~Y.}\ \bibnamefont
  {Shapiro}},\ }\href {\doibase 10.1103/PhysRevB.100.054514} {\bibfield
  {journal} {\bibinfo  {journal} {Phys. Rev. B}\ }\textbf {\bibinfo {volume}
  {100}},\ \bibinfo {pages} {054514} (\bibinfo {year} {2019})}\BibitemShut
  {NoStop}%
\bibitem [{\citenamefont {Phillips}(1989)}]{Phillips}%
  \BibitemOpen
  \bibfield  {author} {\bibinfo {author} {\bibfnamefont {J.}~\bibnamefont
  {Phillips}},\ }\href@noop {} {\emph {\bibinfo {title} {Physics of High-Tc
  Superconductors}}}\ (\bibinfo  {publisher} {Academic Press},\ \bibinfo {year}
  {1989})\ pp.\ \bibinfo {pages} {1--25}\BibitemShut {NoStop}%
\bibitem [{\citenamefont {Rosenstein}\ and\ \citenamefont
  {Shapiro}(2021)}]{Shapiro2021}%
  \BibitemOpen
  \bibfield  {author} {\bibinfo {author} {\bibfnamefont {B.}~\bibnamefont
  {Rosenstein}}\ and\ \bibinfo {author} {\bibfnamefont {B.~Y.}\ \bibnamefont
  {Shapiro}},\ }\href {\doibase 10.1088/2399-6528/abffc3} {\bibfield  {journal}
  {\bibinfo  {journal} {Journal of Physics Communications}\ }\textbf {\bibinfo
  {volume} {5}},\ \bibinfo {pages} {055013} (\bibinfo {year}
  {2021})}\BibitemShut {NoStop}%
\bibitem [{\citenamefont {{Sun}}\ \emph {et~al.}(2020)\citenamefont {{Sun}},
  \citenamefont {{Minkov}}, \citenamefont {{Mozaffari}}, \citenamefont
  {{Chariton}}, \citenamefont {{Prakapenka}}, \citenamefont {{Eremets}},
  \citenamefont {{Balicas}},\ and\ \citenamefont {{Balakirev}}}]{Eremets2020}%
  \BibitemOpen
  \bibfield  {author} {\bibinfo {author} {\bibfnamefont {D.}~\bibnamefont
  {{Sun}}}, \bibinfo {author} {\bibfnamefont {V.~S.}\ \bibnamefont {{Minkov}}},
  \bibinfo {author} {\bibfnamefont {S.}~\bibnamefont {{Mozaffari}}}, \bibinfo
  {author} {\bibfnamefont {S.}~\bibnamefont {{Chariton}}}, \bibinfo {author}
  {\bibfnamefont {V.~B.}\ \bibnamefont {{Prakapenka}}}, \bibinfo {author}
  {\bibfnamefont {M.~I.}\ \bibnamefont {{Eremets}}}, \bibinfo {author}
  {\bibfnamefont {L.}~\bibnamefont {{Balicas}}}, \ and\ \bibinfo {author}
  {\bibfnamefont {F.~F.}\ \bibnamefont {{Balakirev}}},\ }\href@noop {}
  {\bibfield  {journal} {\bibinfo  {journal} {arXiv e-prints}\ ,\ \bibinfo
  {eid} {arXiv:2010.00160}} (\bibinfo {year} {2020})},\ \Eprint
  {http://arxiv.org/abs/2010.00160} {arXiv:2010.00160 [cond-mat.supr-con]}
  \BibitemShut {NoStop}%
\bibitem [{\citenamefont {Chen}(2020)}]{Chen2020}%
  \BibitemOpen
  \bibfield  {author} {\bibinfo {author} {\bibfnamefont {X.-J.}\ \bibnamefont
  {Chen}},\ }\href {\doibase 10.1063/5.0033143} {\bibfield  {journal} {\bibinfo
   {journal} {Matter and Radiation at Extremes}\ }\textbf {\bibinfo {volume}
  {5}},\ \bibinfo {pages} {068102} (\bibinfo {year} {2020})},\ \Eprint
  {http://arxiv.org/abs/https://doi.org/10.1063/5.0033143}
  {https://doi.org/10.1063/5.0033143} \BibitemShut {NoStop}%
\bibitem [{\citenamefont {{Rawal}}\ \emph {et~al.}(2021)\citenamefont
  {{Rawal}}, \citenamefont {{Chang}}, \citenamefont {{Liu}}, \citenamefont
  {{Lu}},\ and\ \citenamefont {{Ting}}}]{Rawal2021}%
  \BibitemOpen
  \bibfield  {author} {\bibinfo {author} {\bibfnamefont {T.~B.}\ \bibnamefont
  {{Rawal}}}, \bibinfo {author} {\bibfnamefont {L.-H.}\ \bibnamefont
  {{Chang}}}, \bibinfo {author} {\bibfnamefont {H.-D.}\ \bibnamefont {{Liu}}},
  \bibinfo {author} {\bibfnamefont {H.-Y.}\ \bibnamefont {{Lu}}}, \ and\
  \bibinfo {author} {\bibfnamefont {C.~S.}\ \bibnamefont {{Ting}}},\
  }\href@noop {} {\bibfield  {journal} {\bibinfo  {journal} {arXiv e-prints}\
  ,\ \bibinfo {eid} {arXiv:2110.00105}} (\bibinfo {year} {2021})},\ \Eprint
  {http://arxiv.org/abs/2110.00105} {arXiv:2110.00105 [cond-mat.supr-con]}
  \BibitemShut {NoStop}%
\bibitem [{\citenamefont {{Sadovskii}}(2021)}]{Sadovskii}%
  \BibitemOpen
  \bibfield  {author} {\bibinfo {author} {\bibfnamefont {M.~V.}\ \bibnamefont
  {{Sadovskii}}},\ }\href@noop {} {\bibfield  {journal} {\bibinfo  {journal}
  {arXiv e-prints}\ ,\ \bibinfo {eid} {arXiv:2106.09948}} (\bibinfo {year}
  {2021})},\ \Eprint {http://arxiv.org/abs/2106.09948} {arXiv:2106.09948
  [cond-mat.supr-con]} \BibitemShut {NoStop}%
\bibitem [{\citenamefont {Testardi}(1975)}]{Testardi}%
  \BibitemOpen
  \bibfield  {author} {\bibinfo {author} {\bibfnamefont {L.~R.}\ \bibnamefont
  {Testardi}},\ }\href {\doibase 10.1103/RevModPhys.47.637} {\bibfield
  {journal} {\bibinfo  {journal} {Rev. Mod. Phys.}\ }\textbf {\bibinfo {volume}
  {47}},\ \bibinfo {pages} {637} (\bibinfo {year} {1975})}\BibitemShut
  {NoStop}%
\bibitem [{\citenamefont {Le~Tacon}\ \emph {et~al.}(2014)\citenamefont
  {Le~Tacon}, \citenamefont {Bosak}, \citenamefont {Souliou}, \citenamefont
  {Dellea}, \citenamefont {Loew}, \citenamefont {Heid}, \citenamefont {Bohnen},
  \citenamefont {Ghiringhelli}, \citenamefont {Krisch},\ and\ \citenamefont
  {Keimer}}]{LeTacon2014}%
  \BibitemOpen
  \bibfield  {author} {\bibinfo {author} {\bibfnamefont {M.}~\bibnamefont
  {Le~Tacon}}, \bibinfo {author} {\bibfnamefont {A.}~\bibnamefont {Bosak}},
  \bibinfo {author} {\bibfnamefont {S.~M.}\ \bibnamefont {Souliou}}, \bibinfo
  {author} {\bibfnamefont {G.}~\bibnamefont {Dellea}}, \bibinfo {author}
  {\bibfnamefont {T.}~\bibnamefont {Loew}}, \bibinfo {author} {\bibfnamefont
  {R.}~\bibnamefont {Heid}}, \bibinfo {author} {\bibfnamefont {K.-P.}\
  \bibnamefont {Bohnen}}, \bibinfo {author} {\bibfnamefont {G.}~\bibnamefont
  {Ghiringhelli}}, \bibinfo {author} {\bibfnamefont {M.}~\bibnamefont
  {Krisch}}, \ and\ \bibinfo {author} {\bibfnamefont {B.}~\bibnamefont
  {Keimer}},\ }\href {\doibase 10.1038/nphys2805} {\bibfield  {journal}
  {\bibinfo  {journal} {Nature Physics}\ }\textbf {\bibinfo {volume} {10}},\
  \bibinfo {pages} {52} (\bibinfo {year} {2014})}\BibitemShut {NoStop}%
\bibitem [{\citenamefont {Uchiyama}\ \emph {et~al.}(2004)\citenamefont
  {Uchiyama}, \citenamefont {Baron}, \citenamefont {Tsutsui}, \citenamefont
  {Tanaka}, \citenamefont {Hu}, \citenamefont {Yamamoto}, \citenamefont
  {Tajima},\ and\ \citenamefont {Endoh}}]{Uchiyama2004}%
  \BibitemOpen
  \bibfield  {author} {\bibinfo {author} {\bibfnamefont {H.}~\bibnamefont
  {Uchiyama}}, \bibinfo {author} {\bibfnamefont {A.~Q.~R.}\ \bibnamefont
  {Baron}}, \bibinfo {author} {\bibfnamefont {S.}~\bibnamefont {Tsutsui}},
  \bibinfo {author} {\bibfnamefont {Y.}~\bibnamefont {Tanaka}}, \bibinfo
  {author} {\bibfnamefont {W.-Z.}\ \bibnamefont {Hu}}, \bibinfo {author}
  {\bibfnamefont {A.}~\bibnamefont {Yamamoto}}, \bibinfo {author}
  {\bibfnamefont {S.}~\bibnamefont {Tajima}}, \ and\ \bibinfo {author}
  {\bibfnamefont {Y.}~\bibnamefont {Endoh}},\ }\href {\doibase
  10.1103/PhysRevLett.92.197005} {\bibfield  {journal} {\bibinfo  {journal}
  {Phys. Rev. Lett.}\ }\textbf {\bibinfo {volume} {92}},\ \bibinfo {pages}
  {197005} (\bibinfo {year} {2004})}\BibitemShut {NoStop}%
\bibitem [{\citenamefont {d'Astuto}\ \emph {et~al.}(2008)\citenamefont
  {d'Astuto}, \citenamefont {Dhalenne}, \citenamefont {Graf}, \citenamefont
  {Hoesch}, \citenamefont {Giura}, \citenamefont {Krisch}, \citenamefont
  {Berthet}, \citenamefont {Lanzara},\ and\ \citenamefont
  {Shukla}}]{Dastuto2008}%
  \BibitemOpen
  \bibfield  {author} {\bibinfo {author} {\bibfnamefont {M.}~\bibnamefont
  {d'Astuto}}, \bibinfo {author} {\bibfnamefont {G.}~\bibnamefont {Dhalenne}},
  \bibinfo {author} {\bibfnamefont {J.}~\bibnamefont {Graf}}, \bibinfo {author}
  {\bibfnamefont {M.}~\bibnamefont {Hoesch}}, \bibinfo {author} {\bibfnamefont
  {P.}~\bibnamefont {Giura}}, \bibinfo {author} {\bibfnamefont
  {M.}~\bibnamefont {Krisch}}, \bibinfo {author} {\bibfnamefont
  {P.}~\bibnamefont {Berthet}}, \bibinfo {author} {\bibfnamefont
  {A.}~\bibnamefont {Lanzara}}, \ and\ \bibinfo {author} {\bibfnamefont
  {A.}~\bibnamefont {Shukla}},\ }\href {\doibase 10.1103/PhysRevB.78.140511}
  {\bibfield  {journal} {\bibinfo  {journal} {Phys. Rev. B}\ }\textbf {\bibinfo
  {volume} {78}},\ \bibinfo {pages} {140511} (\bibinfo {year}
  {2008})}\BibitemShut {NoStop}%
\bibitem [{\citenamefont {Reznik}\ \emph {et~al.}(2008)\citenamefont {Reznik},
  \citenamefont {Fukuda}, \citenamefont {Lamago}, \citenamefont {Baron},
  \citenamefont {Tsutsui}, \citenamefont {Fujita},\ and\ \citenamefont
  {Yamada}}]{Reznik2008}%
  \BibitemOpen
  \bibfield  {author} {\bibinfo {author} {\bibfnamefont {D.}~\bibnamefont
  {Reznik}}, \bibinfo {author} {\bibfnamefont {T.}~\bibnamefont {Fukuda}},
  \bibinfo {author} {\bibfnamefont {D.}~\bibnamefont {Lamago}}, \bibinfo
  {author} {\bibfnamefont {A.}~\bibnamefont {Baron}}, \bibinfo {author}
  {\bibfnamefont {S.}~\bibnamefont {Tsutsui}}, \bibinfo {author} {\bibfnamefont
  {M.}~\bibnamefont {Fujita}}, \ and\ \bibinfo {author} {\bibfnamefont
  {K.}~\bibnamefont {Yamada}},\ }\href {\doibase
  https://doi.org/10.1016/j.jpcs.2008.06.027} {\bibfield  {journal} {\bibinfo
  {journal} {Journal of Physics and Chemistry of Solids}\ }\textbf {\bibinfo
  {volume} {69}},\ \bibinfo {pages} {3103} (\bibinfo {year} {2008})},\ \bibinfo
  {note} {sNS2007}\BibitemShut {NoStop}%
\bibitem [{\citenamefont {Lee}\ \emph {et~al.}(2021)\citenamefont {Lee},
  \citenamefont {Zhou}, \citenamefont {Hepting}, \citenamefont {Li},
  \citenamefont {Nag}, \citenamefont {Walters}, \citenamefont
  {Garcia-Fernandez}, \citenamefont {Robarts}, \citenamefont {Hashimoto},
  \citenamefont {Lu}, \citenamefont {Nosarzewski}, \citenamefont {Song},
  \citenamefont {Eisaki}, \citenamefont {Shen}, \citenamefont {Moritz},
  \citenamefont {Zaanen},\ and\ \citenamefont {Devereaux}}]{Lee2021}%
  \BibitemOpen
  \bibfield  {author} {\bibinfo {author} {\bibfnamefont {W.~S.}\ \bibnamefont
  {Lee}}, \bibinfo {author} {\bibfnamefont {K.-J.}\ \bibnamefont {Zhou}},
  \bibinfo {author} {\bibfnamefont {M.}~\bibnamefont {Hepting}}, \bibinfo
  {author} {\bibfnamefont {J.}~\bibnamefont {Li}}, \bibinfo {author}
  {\bibfnamefont {A.}~\bibnamefont {Nag}}, \bibinfo {author} {\bibfnamefont
  {A.~C.}\ \bibnamefont {Walters}}, \bibinfo {author} {\bibfnamefont
  {M.}~\bibnamefont {Garcia-Fernandez}}, \bibinfo {author} {\bibfnamefont
  {H.~C.}\ \bibnamefont {Robarts}}, \bibinfo {author} {\bibfnamefont
  {M.}~\bibnamefont {Hashimoto}}, \bibinfo {author} {\bibfnamefont
  {H.}~\bibnamefont {Lu}}, \bibinfo {author} {\bibfnamefont {B.}~\bibnamefont
  {Nosarzewski}}, \bibinfo {author} {\bibfnamefont {D.}~\bibnamefont {Song}},
  \bibinfo {author} {\bibfnamefont {H.}~\bibnamefont {Eisaki}}, \bibinfo
  {author} {\bibfnamefont {Z.~X.}\ \bibnamefont {Shen}}, \bibinfo {author}
  {\bibfnamefont {B.}~\bibnamefont {Moritz}}, \bibinfo {author} {\bibfnamefont
  {J.}~\bibnamefont {Zaanen}}, \ and\ \bibinfo {author} {\bibfnamefont {T.~P.}\
  \bibnamefont {Devereaux}},\ }\href {\doibase 10.1038/s41567-020-0993-7}
  {\bibfield  {journal} {\bibinfo  {journal} {Nature Physics}\ }\textbf
  {\bibinfo {volume} {17}},\ \bibinfo {pages} {53} (\bibinfo {year}
  {2021})}\BibitemShut {NoStop}%
\bibitem [{\citenamefont {Sarkar}\ \emph {et~al.}(2021)\citenamefont {Sarkar},
  \citenamefont {Grandadam},\ and\ \citenamefont {P\'epin}}]{Pepin2021}%
  \BibitemOpen
  \bibfield  {author} {\bibinfo {author} {\bibfnamefont {S.}~\bibnamefont
  {Sarkar}}, \bibinfo {author} {\bibfnamefont {M.}~\bibnamefont {Grandadam}}, \
  and\ \bibinfo {author} {\bibfnamefont {C.}~\bibnamefont {P\'epin}},\ }\href
  {\doibase 10.1103/PhysRevResearch.3.013162} {\bibfield  {journal} {\bibinfo
  {journal} {Phys. Rev. Research}\ }\textbf {\bibinfo {volume} {3}},\ \bibinfo
  {pages} {013162} (\bibinfo {year} {2021})}\BibitemShut {NoStop}%
\bibitem [{\citenamefont {Valentinis}\ \emph {et~al.}(2013)\citenamefont
  {Valentinis}, \citenamefont {Berthod}, \citenamefont {Bordini},\ and\
  \citenamefont {Rossi}}]{Valentinis}%
  \BibitemOpen
  \bibfield  {author} {\bibinfo {author} {\bibfnamefont {D.~F.}\ \bibnamefont
  {Valentinis}}, \bibinfo {author} {\bibfnamefont {C.}~\bibnamefont {Berthod}},
  \bibinfo {author} {\bibfnamefont {B.}~\bibnamefont {Bordini}}, \ and\
  \bibinfo {author} {\bibfnamefont {L.}~\bibnamefont {Rossi}},\ }\href
  {\doibase 10.1088/0953-2048/27/2/025008} {\bibfield  {journal} {\bibinfo
  {journal} {Superconductor Science and Technology}\ }\textbf {\bibinfo
  {volume} {27}},\ \bibinfo {pages} {025008} (\bibinfo {year}
  {2013})}\BibitemShut {NoStop}%
\bibitem [{\citenamefont {Setty}\ \emph {et~al.}(2020)\citenamefont {Setty},
  \citenamefont {Baggioli},\ and\ \citenamefont {Zaccone}}]{Setty2020}%
  \BibitemOpen
  \bibfield  {author} {\bibinfo {author} {\bibfnamefont {C.}~\bibnamefont
  {Setty}}, \bibinfo {author} {\bibfnamefont {M.}~\bibnamefont {Baggioli}}, \
  and\ \bibinfo {author} {\bibfnamefont {A.}~\bibnamefont {Zaccone}},\ }\href
  {\doibase 10.1103/PhysRevB.102.174506} {\bibfield  {journal} {\bibinfo
  {journal} {Phys. Rev. B}\ }\textbf {\bibinfo {volume} {102}},\ \bibinfo
  {pages} {174506} (\bibinfo {year} {2020})}\BibitemShut {NoStop}%
\bibitem [{\citenamefont {Setty}\ \emph
  {et~al.}(2021{\natexlab{a}})\citenamefont {Setty}, \citenamefont {Baggioli},\
  and\ \citenamefont {Zaccone}}]{Setty2021}%
  \BibitemOpen
  \bibfield  {author} {\bibinfo {author} {\bibfnamefont {C.}~\bibnamefont
  {Setty}}, \bibinfo {author} {\bibfnamefont {M.}~\bibnamefont {Baggioli}}, \
  and\ \bibinfo {author} {\bibfnamefont {A.}~\bibnamefont {Zaccone}},\ }\href
  {\doibase 10.1103/PhysRevB.103.094519} {\bibfield  {journal} {\bibinfo
  {journal} {Phys. Rev. B}\ }\textbf {\bibinfo {volume} {103}},\ \bibinfo
  {pages} {094519} (\bibinfo {year} {2021}{\natexlab{a}})}\BibitemShut
  {NoStop}%
\bibitem [{\citenamefont {Mizukami}\ \emph {et~al.}(2020)\citenamefont
  {Mizukami}, \citenamefont {Ko\ifmmode~\acute{n}\else \'{n}\fi{}czykowski},
  \citenamefont {Tanaka}, \citenamefont {Juraszek}, \citenamefont {Henkie},
  \citenamefont {Cichorek},\ and\ \citenamefont {Shibauchi}}]{Shibauchi}%
  \BibitemOpen
  \bibfield  {author} {\bibinfo {author} {\bibfnamefont {Y.}~\bibnamefont
  {Mizukami}}, \bibinfo {author} {\bibfnamefont {M.}~\bibnamefont
  {Ko\ifmmode~\acute{n}\else \'{n}\fi{}czykowski}}, \bibinfo {author}
  {\bibfnamefont {O.}~\bibnamefont {Tanaka}}, \bibinfo {author} {\bibfnamefont
  {J.}~\bibnamefont {Juraszek}}, \bibinfo {author} {\bibfnamefont
  {Z.}~\bibnamefont {Henkie}}, \bibinfo {author} {\bibfnamefont
  {T.}~\bibnamefont {Cichorek}}, \ and\ \bibinfo {author} {\bibfnamefont
  {T.}~\bibnamefont {Shibauchi}},\ }\href {\doibase
  10.1103/PhysRevResearch.2.043428} {\bibfield  {journal} {\bibinfo  {journal}
  {Phys. Rev. Research}\ }\textbf {\bibinfo {volume} {2}},\ \bibinfo {pages}
  {043428} (\bibinfo {year} {2020})}\BibitemShut {NoStop}%
\bibitem [{\citenamefont {Kosevich}(2005)}]{Kosevich}%
  \BibitemOpen
  \bibfield  {author} {\bibinfo {author} {\bibfnamefont {A.~M.}\ \bibnamefont
  {Kosevich}},\ }\href@noop {} {\emph {\bibinfo {title} {The Crystal
  Lattice}}}\ (\bibinfo  {publisher} {John Wiley \& Sons, Ltd, Weinheim},\
  \bibinfo {year} {2005})\ pp.\ \bibinfo {pages} {1--13}\BibitemShut {NoStop}%
\bibitem [{\citenamefont {Klemens}(1966{\natexlab{a}})}]{Klemens}%
  \BibitemOpen
  \bibfield  {author} {\bibinfo {author} {\bibfnamefont {P.~G.}\ \bibnamefont
  {Klemens}},\ }\href {\doibase 10.1103/PhysRev.148.845} {\bibfield  {journal}
  {\bibinfo  {journal} {Phys. Rev.}\ }\textbf {\bibinfo {volume} {148}},\
  \bibinfo {pages} {845} (\bibinfo {year} {1966}{\natexlab{a}})}\BibitemShut
  {NoStop}%
\bibitem [{\citenamefont {Kittel}(2005)}]{Kittel}%
  \BibitemOpen
  \bibfield  {author} {\bibinfo {author} {\bibfnamefont {C.}~\bibnamefont
  {Kittel}},\ }\href@noop {} {\emph {\bibinfo {title} {Introduction to solid
  state physics. Eighth edition}}}\ (\bibinfo  {publisher} {John Wiley \&
  Sons},\ \bibinfo {year} {2005})\BibitemShut {NoStop}%
\bibitem [{\citenamefont {Gor{\textquoteright}kov}(2016)}]{Gorkov}%
  \BibitemOpen
  \bibfield  {author} {\bibinfo {author} {\bibfnamefont {L.~P.}\ \bibnamefont
  {Gor{\textquoteright}kov}},\ }\href {\doibase 10.1073/pnas.1604145113}
  {\bibfield  {journal} {\bibinfo  {journal} {Proceedings of the National
  Academy of Sciences}\ }\textbf {\bibinfo {volume} {113}},\ \bibinfo {pages}
  {4646} (\bibinfo {year} {2016})}\BibitemShut {NoStop}%
\bibitem [{\citenamefont {Gastiasoro}\ \emph
  {et~al.}(2020{\natexlab{b}})\citenamefont {Gastiasoro}, \citenamefont
  {Trevisan},\ and\ \citenamefont {Fernandes}}]{Gastiasoro2020}%
  \BibitemOpen
  \bibfield  {author} {\bibinfo {author} {\bibfnamefont {M.~N.}\ \bibnamefont
  {Gastiasoro}}, \bibinfo {author} {\bibfnamefont {T.~V.}\ \bibnamefont
  {Trevisan}}, \ and\ \bibinfo {author} {\bibfnamefont {R.~M.}\ \bibnamefont
  {Fernandes}},\ }\href {\doibase 10.1103/PhysRevB.101.174501} {\bibfield
  {journal} {\bibinfo  {journal} {Phys. Rev. B}\ }\textbf {\bibinfo {volume}
  {101}},\ \bibinfo {pages} {174501} (\bibinfo {year}
  {2020}{\natexlab{b}})}\BibitemShut {NoStop}%
\bibitem [{\citenamefont {{Gastiasoro}}\ \emph {et~al.}(2021)\citenamefont
  {{Gastiasoro}}, \citenamefont {{Eleonora Temperini}}, \citenamefont
  {{Barone}},\ and\ \citenamefont {{Lorenzana}}}]{Gastiasoro2021}%
  \BibitemOpen
  \bibfield  {author} {\bibinfo {author} {\bibfnamefont {M.~N.}\ \bibnamefont
  {{Gastiasoro}}}, \bibinfo {author} {\bibfnamefont {M.}~\bibnamefont
  {{Eleonora Temperini}}}, \bibinfo {author} {\bibfnamefont {P.}~\bibnamefont
  {{Barone}}}, \ and\ \bibinfo {author} {\bibfnamefont {J.}~\bibnamefont
  {{Lorenzana}}},\ }\href@noop {} {\bibfield  {journal} {\bibinfo  {journal}
  {arXiv e-prints}\ ,\ \bibinfo {eid} {arXiv:2109.13207}} (\bibinfo {year}
  {2021})},\ \Eprint {http://arxiv.org/abs/2109.13207} {arXiv:2109.13207
  [cond-mat.supr-con]} \BibitemShut {NoStop}%
\bibitem [{\citenamefont {van~der Marel}\ \emph {et~al.}(2019)\citenamefont
  {van~der Marel}, \citenamefont {Barantani},\ and\ \citenamefont
  {Rischau}}]{Rischau}%
  \BibitemOpen
  \bibfield  {author} {\bibinfo {author} {\bibfnamefont {D.}~\bibnamefont
  {van~der Marel}}, \bibinfo {author} {\bibfnamefont {F.}~\bibnamefont
  {Barantani}}, \ and\ \bibinfo {author} {\bibfnamefont {C.~W.}\ \bibnamefont
  {Rischau}},\ }\href {\doibase 10.1103/PhysRevResearch.1.013003} {\bibfield
  {journal} {\bibinfo  {journal} {Phys. Rev. Research}\ }\textbf {\bibinfo
  {volume} {1}},\ \bibinfo {pages} {013003} (\bibinfo {year}
  {2019})}\BibitemShut {NoStop}%
\bibitem [{\citenamefont {{Volkov}}\ \emph {et~al.}(2021)\citenamefont
  {{Volkov}}, \citenamefont {{Chandra}},\ and\ \citenamefont
  {{Coleman}}}]{Coleman2021}%
  \BibitemOpen
  \bibfield  {author} {\bibinfo {author} {\bibfnamefont {P.~A.}\ \bibnamefont
  {{Volkov}}}, \bibinfo {author} {\bibfnamefont {P.}~\bibnamefont {{Chandra}}},
  \ and\ \bibinfo {author} {\bibfnamefont {P.}~\bibnamefont {{Coleman}}},\
  }\href@noop {} {\bibfield  {journal} {\bibinfo  {journal} {arXiv e-prints}\
  ,\ \bibinfo {eid} {arXiv:2106.11295}} (\bibinfo {year} {2021})},\ \Eprint
  {http://arxiv.org/abs/2106.11295} {arXiv:2106.11295 [cond-mat.supr-con]}
  \BibitemShut {NoStop}%
\bibitem [{\citenamefont {{Kiseliov}}\ and\ \citenamefont
  {{Feigel'man}}(2021)}]{Feigelman}%
  \BibitemOpen
  \bibfield  {author} {\bibinfo {author} {\bibfnamefont {D.}~\bibnamefont
  {{Kiseliov}}}\ and\ \bibinfo {author} {\bibfnamefont {M.}~\bibnamefont
  {{Feigel'man}}},\ }\href@noop {} {\bibfield  {journal} {\bibinfo  {journal}
  {arXiv e-prints}\ ,\ \bibinfo {eid} {arXiv:2106.09530}} (\bibinfo {year}
  {2021})},\ \Eprint {http://arxiv.org/abs/2106.09530} {arXiv:2106.09530
  [cond-mat.supr-con]} \BibitemShut {NoStop}%
\bibitem [{\citenamefont {Ngai}(1974)}]{Ngai}%
  \BibitemOpen
  \bibfield  {author} {\bibinfo {author} {\bibfnamefont {K.~L.}\ \bibnamefont
  {Ngai}},\ }\href {\doibase 10.1103/PhysRevLett.32.215} {\bibfield  {journal}
  {\bibinfo  {journal} {Phys. Rev. Lett.}\ }\textbf {\bibinfo {volume} {32}},\
  \bibinfo {pages} {215} (\bibinfo {year} {1974})}\BibitemShut {NoStop}%
\bibitem [{\citenamefont {Xu}\ \emph {et~al.}(2008)\citenamefont {Xu},
  \citenamefont {Wang}, \citenamefont {Duan}, \citenamefont {Gu},\ and\
  \citenamefont {Li}}]{Baowen}%
  \BibitemOpen
  \bibfield  {author} {\bibinfo {author} {\bibfnamefont {Y.}~\bibnamefont
  {Xu}}, \bibinfo {author} {\bibfnamefont {J.-S.}\ \bibnamefont {Wang}},
  \bibinfo {author} {\bibfnamefont {W.}~\bibnamefont {Duan}}, \bibinfo {author}
  {\bibfnamefont {B.-L.}\ \bibnamefont {Gu}}, \ and\ \bibinfo {author}
  {\bibfnamefont {B.}~\bibnamefont {Li}},\ }\href {\doibase
  10.1103/PhysRevB.78.224303} {\bibfield  {journal} {\bibinfo  {journal} {Phys.
  Rev. B}\ }\textbf {\bibinfo {volume} {78}},\ \bibinfo {pages} {224303}
  (\bibinfo {year} {2008})}\BibitemShut {NoStop}%
\bibitem [{\citenamefont {Tadano}\ and\ \citenamefont
  {Tsuneyuki}(2018)}]{Tadano1}%
  \BibitemOpen
  \bibfield  {author} {\bibinfo {author} {\bibfnamefont {T.}~\bibnamefont
  {Tadano}}\ and\ \bibinfo {author} {\bibfnamefont {S.}~\bibnamefont
  {Tsuneyuki}},\ }\href {\doibase 10.7566/JPSJ.87.041015} {\bibfield  {journal}
  {\bibinfo  {journal} {Journal of the Physical Society of Japan}\ }\textbf
  {\bibinfo {volume} {87}},\ \bibinfo {pages} {041015} (\bibinfo {year}
  {2018})},\ \Eprint
  {http://arxiv.org/abs/https://doi.org/10.7566/JPSJ.87.041015}
  {https://doi.org/10.7566/JPSJ.87.041015} \BibitemShut {NoStop}%
\bibitem [{\citenamefont {Tadano}\ and\ \citenamefont
  {Tsuneyuki}(2015)}]{Tadano2}%
  \BibitemOpen
  \bibfield  {author} {\bibinfo {author} {\bibfnamefont {T.}~\bibnamefont
  {Tadano}}\ and\ \bibinfo {author} {\bibfnamefont {S.}~\bibnamefont
  {Tsuneyuki}},\ }\href {\doibase 10.1103/PhysRevB.92.054301} {\bibfield
  {journal} {\bibinfo  {journal} {Phys. Rev. B}\ }\textbf {\bibinfo {volume}
  {92}},\ \bibinfo {pages} {054301} (\bibinfo {year} {2015})}\BibitemShut
  {NoStop}%
\bibitem [{\citenamefont {Li}\ \emph {et~al.}(2015)\citenamefont {Li},
  \citenamefont {Hong}, \citenamefont {May}, \citenamefont {Bansal},
  \citenamefont {Chi}, \citenamefont {Hong}, \citenamefont {Ehlers},\ and\
  \citenamefont {Delaire}}]{Delaire2015}%
  \BibitemOpen
  \bibfield  {author} {\bibinfo {author} {\bibfnamefont {C.~W.}\ \bibnamefont
  {Li}}, \bibinfo {author} {\bibfnamefont {J.}~\bibnamefont {Hong}}, \bibinfo
  {author} {\bibfnamefont {A.~F.}\ \bibnamefont {May}}, \bibinfo {author}
  {\bibfnamefont {D.}~\bibnamefont {Bansal}}, \bibinfo {author} {\bibfnamefont
  {S.}~\bibnamefont {Chi}}, \bibinfo {author} {\bibfnamefont {T.}~\bibnamefont
  {Hong}}, \bibinfo {author} {\bibfnamefont {G.}~\bibnamefont {Ehlers}}, \ and\
  \bibinfo {author} {\bibfnamefont {O.}~\bibnamefont {Delaire}},\ }\href
  {\doibase 10.1038/nphys3492} {\bibfield  {journal} {\bibinfo  {journal}
  {Nature Physics}\ }\textbf {\bibinfo {volume} {11}},\ \bibinfo {pages} {1063}
  (\bibinfo {year} {2015})}\BibitemShut {NoStop}%
\bibitem [{\citenamefont {Casella}\ and\ \citenamefont
  {Zaccone}(2021)}]{Casella}%
  \BibitemOpen
  \bibfield  {author} {\bibinfo {author} {\bibfnamefont {L.}~\bibnamefont
  {Casella}}\ and\ \bibinfo {author} {\bibfnamefont {A.}~\bibnamefont
  {Zaccone}},\ }\href {\doibase 10.1088/1361-648x/abdb68} {\bibfield  {journal}
  {\bibinfo  {journal} {Journal of Physics: Condensed Matter}\ }\textbf
  {\bibinfo {volume} {33}},\ \bibinfo {pages} {165401} (\bibinfo {year}
  {2021})}\BibitemShut {NoStop}%
\bibitem [{\citenamefont {Setty}\ \emph
  {et~al.}(2021{\natexlab{b}})\citenamefont {Setty}, \citenamefont
  {Fanfarillo},\ and\ \citenamefont {Hirschfeld}}]{Setty2021-FPDW}%
  \BibitemOpen
  \bibfield  {author} {\bibinfo {author} {\bibfnamefont {C.}~\bibnamefont
  {Setty}}, \bibinfo {author} {\bibfnamefont {L.}~\bibnamefont {Fanfarillo}}, \
  and\ \bibinfo {author} {\bibfnamefont {P.}~\bibnamefont {Hirschfeld}},\
  }\href@noop {} {\bibfield  {journal} {\bibinfo  {journal} {arXiv preprint
  arXiv:2110.13138}\ } (\bibinfo {year} {2021}{\natexlab{b}})}\BibitemShut
  {NoStop}%
\bibitem [{\citenamefont {Marsiglio}\ and\ \citenamefont
  {Carbotte}(2008)}]{Marsiglio}%
  \BibitemOpen
  \bibfield  {author} {\bibinfo {author} {\bibfnamefont {F.}~\bibnamefont
  {Marsiglio}}\ and\ \bibinfo {author} {\bibfnamefont {J.}~\bibnamefont
  {Carbotte}},\ }in\ \href@noop {} {\emph {\bibinfo {booktitle}
  {Superconductivity}}}\ (\bibinfo  {publisher} {Springer},\ \bibinfo {year}
  {2008})\ pp.\ \bibinfo {pages} {73--162}\BibitemShut {NoStop}%
\bibitem [{\citenamefont {Kleinert}(2018)}]{Kleinert}%
  \BibitemOpen
  \bibfield  {author} {\bibinfo {author} {\bibfnamefont {H.}~\bibnamefont
  {Kleinert}},\ }\href {\doibase 10.1142/10545} {\emph {\bibinfo {title}
  {Collective classical and quantum fields}}}\ (\bibinfo  {publisher} {World
  Scientific, Singapore},\ \bibinfo {year} {2018})\BibitemShut {NoStop}%
\bibitem [{\citenamefont {Ziman}(1969)}]{Ziman}%
  \BibitemOpen
  \bibfield  {author} {\bibinfo {author} {\bibfnamefont {J.~M.}\ \bibnamefont
  {Ziman}},\ }\href@noop {} {\emph {\bibinfo {title} {Elements of Advanced
  Quantum Theory}}}\ (\bibinfo  {publisher} {Cambridge University Press,
  Cambridge},\ \bibinfo {year} {1969})\BibitemShut {NoStop}%
\bibitem [{\citenamefont {Klemens}(1966{\natexlab{b}})}]{Klemens1966}%
  \BibitemOpen
  \bibfield  {author} {\bibinfo {author} {\bibfnamefont {P.}~\bibnamefont
  {Klemens}},\ }\href@noop {} {\bibfield  {journal} {\bibinfo  {journal}
  {Physical Review}\ }\textbf {\bibinfo {volume} {148}},\ \bibinfo {pages}
  {845} (\bibinfo {year} {1966}{\natexlab{b}})}\BibitemShut {NoStop}%
\bibitem [{\citenamefont {Larkin}\ and\ \citenamefont
  {Varlamov}(2005)}]{Varlamov2005}%
  \BibitemOpen
  \bibfield  {author} {\bibinfo {author} {\bibfnamefont {A.}~\bibnamefont
  {Larkin}}\ and\ \bibinfo {author} {\bibfnamefont {A.}~\bibnamefont
  {Varlamov}},\ }\href@noop {} {\emph {\bibinfo {title} {Theory of fluctuations
  in superconductors}}}\ (\bibinfo  {publisher} {Clarendon Press},\ \bibinfo
  {year} {2005})\BibitemShut {NoStop}%
\bibitem [{\citenamefont {Lanigan-Atkins}\ \emph {et~al.}(2020)\citenamefont
  {Lanigan-Atkins}, \citenamefont {Yang}, \citenamefont {Niedziela},
  \citenamefont {Bansal}, \citenamefont {May}, \citenamefont {Puretzky},
  \citenamefont {Lin}, \citenamefont {Pajerowski}, \citenamefont {Hong},
  \citenamefont {Chi}, \citenamefont {Ehlers},\ and\ \citenamefont
  {Delaire}}]{Delaire2020}%
  \BibitemOpen
  \bibfield  {author} {\bibinfo {author} {\bibfnamefont {T.}~\bibnamefont
  {Lanigan-Atkins}}, \bibinfo {author} {\bibfnamefont {S.}~\bibnamefont
  {Yang}}, \bibinfo {author} {\bibfnamefont {J.~L.}\ \bibnamefont {Niedziela}},
  \bibinfo {author} {\bibfnamefont {D.}~\bibnamefont {Bansal}}, \bibinfo
  {author} {\bibfnamefont {A.~F.}\ \bibnamefont {May}}, \bibinfo {author}
  {\bibfnamefont {A.~A.}\ \bibnamefont {Puretzky}}, \bibinfo {author}
  {\bibfnamefont {J.~Y.~Y.}\ \bibnamefont {Lin}}, \bibinfo {author}
  {\bibfnamefont {D.~M.}\ \bibnamefont {Pajerowski}}, \bibinfo {author}
  {\bibfnamefont {T.}~\bibnamefont {Hong}}, \bibinfo {author} {\bibfnamefont
  {S.}~\bibnamefont {Chi}}, \bibinfo {author} {\bibfnamefont {G.}~\bibnamefont
  {Ehlers}}, \ and\ \bibinfo {author} {\bibfnamefont {O.}~\bibnamefont
  {Delaire}},\ }\href {\doibase 10.1038/s41467-020-18121-4} {\bibfield
  {journal} {\bibinfo  {journal} {Nature Communications}\ }\textbf {\bibinfo
  {volume} {11}},\ \bibinfo {pages} {4430} (\bibinfo {year}
  {2020})}\BibitemShut {NoStop}%
\bibitem [{\citenamefont {Setty}(2019)}]{Setty2019}%
  \BibitemOpen
  \bibfield  {author} {\bibinfo {author} {\bibfnamefont {C.}~\bibnamefont
  {Setty}},\ }\href {\doibase 10.1103/PhysRevB.99.144523} {\bibfield  {journal}
  {\bibinfo  {journal} {Phys. Rev. B}\ }\textbf {\bibinfo {volume} {99}},\
  \bibinfo {pages} {144523} (\bibinfo {year} {2019})}\BibitemShut {NoStop}%
\bibitem [{\citenamefont {Himmetoglu}\ \emph {et~al.}(2014)\citenamefont
  {Himmetoglu}, \citenamefont {Janotti}, \citenamefont {Peelaers},
  \citenamefont {Alkauskas},\ and\ \citenamefont {Van~de Walle}}]{Himmetoglu}%
  \BibitemOpen
  \bibfield  {author} {\bibinfo {author} {\bibfnamefont {B.}~\bibnamefont
  {Himmetoglu}}, \bibinfo {author} {\bibfnamefont {A.}~\bibnamefont {Janotti}},
  \bibinfo {author} {\bibfnamefont {H.}~\bibnamefont {Peelaers}}, \bibinfo
  {author} {\bibfnamefont {A.}~\bibnamefont {Alkauskas}}, \ and\ \bibinfo
  {author} {\bibfnamefont {C.~G.}\ \bibnamefont {Van~de Walle}},\ }\href
  {\doibase 10.1103/PhysRevB.90.241204} {\bibfield  {journal} {\bibinfo
  {journal} {Phys. Rev. B}\ }\textbf {\bibinfo {volume} {90}},\ \bibinfo
  {pages} {241204} (\bibinfo {year} {2014})}\BibitemShut {NoStop}%
\end{thebibliography}%

\end{document}